\documentclass[aps,twocolumn,superscriptaddress,prb]{revtex4-2}
\usepackage{amsfonts}
\usepackage{comment}
\usepackage{amsmath,amssymb,epsfig,color}
\usepackage{float}
\usepackage{amsmath}
\usepackage{amssymb}
\usepackage{graphicx}
\usepackage{dcolumn}
\usepackage{bbold}
\usepackage{physics}
\usepackage{siunitx}
\AtBeginDocument{\RenewCommandCopy\qty\SI} 
\usepackage{hyperref}
\footnotetext{I. B. and L. A. contributed equally to this work.}

\begin{document}

\title{Superconducting diode effect in Ising superconductors}

\author{Itai Bankier\(^*\)}
\affiliation{The Racah Institute of Physics, The Hebrew University of Jerusalem, Jerusalem 91904, Israel}

\author{Lotan Attias\(^*\)}
\affiliation{The Racah Institute of Physics, The Hebrew University of Jerusalem, Jerusalem 91904, Israel}

\author{Alex Levchenko}
\affiliation{Department of Physics, University of Wisconsin--Madison, Madison, Wisconsin 53706, USA}

\author{Maxim Khodas}
\affiliation{The Racah Institute of Physics, The Hebrew University of Jerusalem, Jerusalem 91904, Israel}

\begin{abstract}
We study the superconducting diode effect (SDE) in an Ising superconductor with broken basal mirror symmetry in a parallel magnetic field.  
We show that in the presence of a small Rashba spin splitting, \( \Delta_\text{R} \), the dominant Ising spin-orbit coupling (\( \Delta_\text{I} \gg \Delta_\text{R} \)) dramatically enhances the SDE efficiency compared to a Rashba superconductor with \( \Delta_\text{I} = 0 \) and the same \( \Delta_\text{R} \).   
The suppression of the SDE for \( \Delta_\text{I} = 0 \) at \( \Delta_\text{R} \) much larger than the critical temperature (\( T_\text{c} \)) is accidental.  
At \( \Delta_\text{R} \ll T_\text{c} \), the SDE efficiency is small because, to linear order, \( \Delta_\text{R} \) can be removed by a gauge transformation. 
These two factors -- the accidental suppression of the SDE at large \( \Delta_\text{R} \) and the systematic suppression at small \( \Delta_\text{R} \) -- are eliminated by Ising spin-orbit coupling.  
As a result, SDE efficiency is substantially enhanced for \( \Delta_\text{I} \gg \Delta_\text{R} \neq 0 \).
\end{abstract}
%\date{\today}    
\maketitle
%#####################################################################
%#####################################################################
%#####################################################################
%#####################################################################
%#####################################################################
\section{Introduction}
\label{sec:Intro}

The superconducting diode effect (SDE) signifies the nonreciprocity of the critical current \cite{jiang_superconducting_2022,Nadeem2023}.
The SDE appears when all the unitary and anti-unitary symmetries that reverse the current (\(\mathbf{J}\)) are broken \cite{ZinklHamamotoSigrist22}.
The most common of these are the time reversal ($\mathcal{T}$) and spatial inversion ($\mathcal{P}$) symmetries.
In fact, the SDE was initially considered theoretically in noncentrosymmetric superconductors under an applied magnetic field \cite{Levitov1985,Edelstein1996}.

Interest in the SDE was renewed following its observation in [Nb/V/Ta]$_n$ superlattices \cite{Ando2020}.
In subsequent years, the SDE has been reported in Josephson junctions \cite{BaumgartnerManfraStrunk22,LotfizadehShabani24,Erin2024,Debnath2024} and hybrid multilayer geometries \cite{Miyasaka2021,KarabassovBobkovaVasenko22, SundareshRokhinson23,KealhoferBalentsStemmer23,GengBergeretHeikkila23,SatchellBurnell23,PutilovBuzdin24}.
Furthermore, supercurrent rectification due to the anisotropy of the vortex pinning potential -- such as the vortex ratchet effect -- has been observed \cite{VodolazovPeeters05,CerbuVondel13,GutfreundAnahory23,HouMoodera23,LyuKwok21}.

The SDE is commonly quantified by the difference in critical currents for opposite current polarities, \(J_{\text{c}\pm}\), using the efficiency parameter
\( \eta_{\mathrm{SDE}} = (J_{\text{c}+} + J_{\text{c}-})/(J_{\text{c}+} - J_{\text{c}-}) \). 
A key objective of current research is to identify setups that allow optimal control over SDE efficiency.

A paradigmatic example of a system exhibiting the SDE is the Rashba superconductor \cite{WakatsukiNagaosa18,HoshinoNagaosa18,DaidoYanase22,DaidoYanase22_2,YuanFu22,HeNagaosa22,IlicBergeret22}.  
This system describes a two-dimensional superconductor confined to the \(xy\)-plane, where basal plane mirror symmetry (\( m_{\hat{z}} \)) is broken, and a parallel magnetic field \( \mathbf{B} \perp \hat{z} \) is applied.  
Here, \( m_{\hat{n}} \) denotes mirror reflection in the plane perpendicular to a vector \( \mathbf{n} \).  
The broken \( m_{\hat{z}} \) symmetry leads to a finite band spin splitting of \( 2\Delta_\text{R} \) due to Rashba spin-orbit coupling (SOC) \cite{Rashba1984}.  
As a consequence of the remaining \( m_{\hat{B}} \) symmetry, the SDE efficiency \( \eta_{\mathrm{SDE}} \) vanishes when the current flows parallel to the magnetic field (\( \mathbf{J} \parallel \mathbf{B} \)) and reaches its maximum when the current is perpendicular to the field (\( \mathbf{J} \perp \mathbf{B} \)).

\begin{figure}[t!]
\begin{center}
\centering
\includegraphics[width=0.45\textwidth]{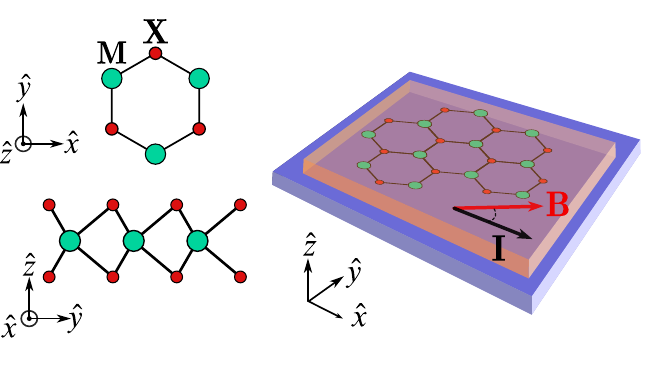}
\caption{ 
Ising superconductor confined to the $xy$-plane on a substrate.
Top and side view of a hexagonal lattice of a \( \text{MX}_2 \) monolayer, where \(\text{M}\) large circles (green) is a transition metal and \(\text{X}\) small circles (red) is a chalcogen. 
The substrate breaks $m_{\hat{z}}$ symmetry introducing the Rashba SOC. 
Current $\mathbf{I}$ flows in the $xy$-plane, and an external magnetic field $\mathbf{B}$ is applied to the system.
} \label{fig:system}
\end{center}
\end{figure}

\begin{figure*}[t!]
\begin{center}
\centering
\includegraphics[width=0.96\textwidth]{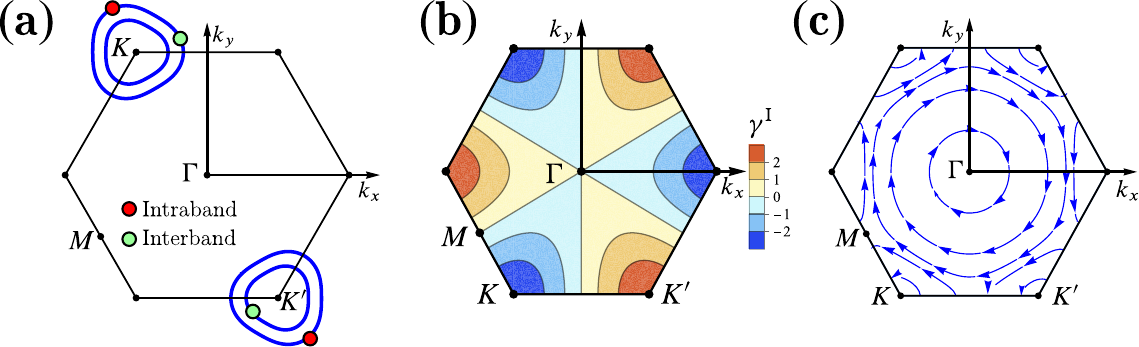}
\caption{
First Brillouin zone of a  transition-metal dichalcogenide MX$_2$ monolayer.
(a) Fermi pockets in $K$ ($K'$)-model of an Ising superconductor at finite trigonal warping, Eq.~\eqref{eq:H_0_warp}. 
Cooper pairs at the outer or inner Fermi pockets are intraband (dark red circles).
When one of the electrons comprising a pair is at the outer (inner) pocket, while the other is at the inner (outer) pocket the Cooper pair is interband (light green circles).
The interband pairing is induced by the interaction Eq.~\eqref{eq:H_p} at finite supercurrent and magnetic field. These perturbations also affect the intraband pairing.
(b) Ising SOC \( \gamma^\text{I} \) (arbitrary units). 
(c) Streamlines of the Rashba SOC, \( \boldsymbol{\gamma}^\text{R}(\mathbf{k}\eta) \).
The SOC presented is based on \cite{Haim2022}.}
\label{fig:model}
\end{center}
\end{figure*}

In this work, we propose the Ising superconductor on a substrate subject to an in-plane magnetic field as a viable alternative to Rashba superconductors (see Fig.~\ref{fig:system}).  
In a pristine Ising superconductor, the Ising SOC polarizes spins out-of-plane (Fig.~\ref{fig:model}b) and coexists with superconductivity \cite{lu2015evidence,saito2016superconductivity,xi2016Ising,costanzo2018tunnelling,delaBarrera2018}.  
The substrate breaks the \( m_{\hat{z}} \) symmetry, leading to a finite Rashba spin splitting (Fig.~\ref{fig:model}c) in addition to the Ising SOC.  
Our main finding is that for a given \( \Delta_\text{R} \neq 0 \), the introduction of Ising SOC significantly enhances the SDE efficiency compared to a Rashba superconductor.  
In transition metal (M) dichalcogenide (X) monolayers, MX$_2$, this amplification can realistically reach two orders of magnitude.

The enhancement of the SDE in Ising superconductors is somewhat counterintuitive because (1) Ising SOC strongly counteracts the in-plane field, and (2) it tends to restore the \( m_{\hat{z}} \) symmetry, which ordinarily prohibits the SDE.  
To understand why, contrary to these expectations, Ising SOC strongly enhances the SDE, we revisit the dependence of the SDE on SOC in the Rashba model.  
We analyze this separately in the limits of strong (\( \kappa_\text{R} \gg 1 \)) and weak (\( \kappa_\text{R} \ll 1 \)) Rashba SOC, where \( \kappa_\text{R} = \Delta_\text{R} /\pi T_\text{c} \), and \( T_\text{c} \) is the superconducting transition temperature.  
In what follows, we restrict ourselves to the Ginzburg-Landau (GL) regime of temperatures (\(T\)) close to \(T_\text{c}\).  
Generally, in the GL formalism, the SDE scales as \( \eta \sim B \sqrt{T_\text{c} - T} \) \cite{HoshinoNagaosa18,DaidoYanase22,DaidoYanase22_2}.  

At strong Rashba SOC, \( \kappa_\text{R} \gg 1 \), the two-band quasi-classical approach \cite{Houzet2015} predicts a finite SDE \cite{IlicBergeret22}.  
However, in the GL regime, it yields a vanishing result at leading order in \( B \sqrt{T_\text{c} - T} \).  
This vanishing SDE arises because of an accidental cancellation of various contributions to the critical current.  

%@@@@@@@@@@@@@@@@@@@@@@@@@@@@@@@@@@@@@@@@@@@@@@@@@@@@@@@@@@@@@@@@@@@@@@@@@@@
%@@@@@@@@@@@@@@@@@@@@@@@@@@@@@@@@@@@@@@@@@@@@@@@@@@@@@@@@@@@@@@@@@@@@@@@@@@@
%@@@@@@@@@@@@@@@@@@@@@@@@@@@@@@@@@@@@@@@@@@@@@@@@@@@@@@@@@@@@@@@@@@@@@@@@@@@
%@@@@@@@@@@@@@@@@@@@@@@@@@@@@@@@@@@@@@@@@@@@@@@@@@@@@@@@@@@@@@@@@@@@@@@@@@@@
%@@@@@@@@@@@@@@@@@@@@@@@@@@@@@@@@@@@@@@@@@@@@@@@@@@@@@@@@@@@@@@@@@@@@@@@@@@@
%@@@@@@@@@@@@@@@@@@@@@@@@@@@@@@@@@@@@@@@@@@@@@@@@@@@@@@@@@@@@@@@@@@@@@@@@@@@

%@@@@@@@@@@@@@@@@@@@@@@@@@@@@@@@@@@@@@@@@@@@@@@@@@@@@@@@@@@@@@@@@@@@@@@@@@@@
%@@@@@@@@@@@@@@@@@@@@@@@@@@@@@@@@@@@@@@@@@@@@@@@@@@@@@@@@@@@@@@@@@@@@@@@@@@@
%@@@@@@@@@@@@@@@@@@@@@@@@@@@@@@@@@@@@@@@@@@@@@@@@@@@@@@@@@@@@@@@@@@@@@@@@@@@
%@@@@@@@@@@@@@@@@@@@@@@@@@@@@@@@@@@@@@@@@@@@@@@@@@@@@@@@@@@@@@@@@@@@@@@@@@@@
%@@@@@@@@@@@@@@@@@@@@@@@@@@@@@@@@@@@@@@@@@@@@@@@@@@@@@@@@@@@@@@@@@@@@@@@@@@@
%@@@@@@@@@@@@@@@@@@@@@@@@@@@@@@@@@@@@@@@@@@@@@@@@@@@@@@@@@@@@@@@@@@@@@@@@@@@

In the opposite limit, \( \kappa_\text{R} \ll 1 \), the SDE vanishes to linear order in Rashba SOC \cite{Hasan2024}.  
Unlike the suppression in the strong Rashba limit, this vanishing SDE is not accidental but follows directly from gauge invariance.  
To see this, recall that Rashba SOC can be interpreted as a coupling to a spin-dependent gauge field \cite{Hatano2007,Tokatly2008}.  
In one-dimensional systems, this gauge field can be eliminated via a spin-dependent gauge transformation \cite{Kaplan1983},  
resulting in an Aharonov-Casher phase \cite{Avishai2019}.  
However, in two dimensions, this transformation is generally not possible, as the spin-orbit gauge field is non-Abelian \cite{Chen2008}, except in special cases \cite{Tokatly2010}.

Nevertheless, to first order in \( \Delta_\text{R} \), the Rashba SOC can be eliminated using a specific gauge transformation \cite{Hatano2007}.  
We argue that this transformation does not affect the local SU(2)-invariant spin-singlet pairing interaction.  
However, it modifies the in-plane magnetic field, effectively converting it into a spatially varying out-of-plane field,  
\( \tilde{\mathbf{B}} \propto \Delta_\text{R} \hat{z}(\mathbf{B} \cdot \mathbf{r}) \), which remains coupled exclusively to spins via the Zeeman interaction.

To first order in \( \Delta_\text{R} \), the Rashba superconductor problem is thus gauge -- equivalent to the problem of magnetization gradients studied in Ref.~\cite{KotetesAndersen23}.  
In particular, it has been shown that magnetization gradients exhibit the same transformation properties as in-plane magnetic fields  
and can be used as an alternative mechanism to generate the SDE \cite{RoigAndersen24}.

Crucially, a nonzero SOC is still required for a finite SDE in the presence of either magnetization gradients or an in-plane magnetic field.  
At zero SOC, the spin-independent current \( \mathbf{J} \perp \mathbf{B} \) is reversed by the spatial mirror symmetry operation \( m'_{\hat{J}} \),  
which acts on the orbital degrees of freedom without affecting spins.  
Since the SOC has been eliminated to first order, this explains the absence of an SDE in the Rashba model at this order.  
On a technical level, this result holds separately for all contributions to the critical current and to all orders in the magnetic field.

Ising SOC transforms nontrivially under the above gauge transformation, thereby invalidating the previous arguments based on gauge invariance.  
In doing so, it prevents both the systematic and accidental suppression of the SDE.  
This is the underlying physical reason for the substantial enhancement of SDE efficiency due to Ising SOC, as studied in this work.

The paper is organized as follows.  
In Sec.~\ref{sec:Model}, we introduce the model and present the results for the SDE in Sec.~\ref{sec:SDE}.  
We then discuss our findings in the context of existing SDE measurements in Sec.~\ref{sec:Discussion} and conclude in Sec.~\ref{sec:Conclusions}.
%%%%%%%%%%%%%%%%%%%%%%%%%%%%%%%%%%%%%%%%%%%%%%%%%%%%%%%%%%%%%%%%%%%%%%%%%%%%%%%%
\section{Model formulation}
\label{sec:Model}

To illustrate the amplification of the SDE by Ising SOC, we focus on a specific model of NbSe$_2$ monolayer.
We stress that qualitatively, our conclusions regarding the enhancement of SDE efficiency are entirely general for any MX$_2$ transition-metal dichalcogenide with  $D_{3h}$ symmetry. 

In NbSe$_2$ bands that cross the Fermi level form the hole pockets centered at $\Gamma$ and $K$ ($K'$) points in the hexagonal Brillouin zone. 
Here, for simplicity, we only consider the pockets centered at $K$ ($K'$), { see Fig.~\ref{fig:model}}.
The Hamiltonian \( H = H_K + V_\text{p} \) contains the single-band $K$ ($K'$) pocket Hamiltonian $H_K$ of non-interacting electrons, and the Cooper pairing Hamiltonian $V_\text{p}$. 

We introduce the operators $c^{\dagger}_{\mathbf{k}\eta s}$ that create the electrons with momentum $\mathbf{k}$ in the pocket $K$ ($K'$) distinguished by the pocket subscript $\eta=+1$ and $\eta=-1$, respectively, with spin $s$ along $\hat{z}$.
Here the momentum is counted relative to the $K$ ($K'$) as determined by the subscript $\eta$.
We write $H_K = \sum_{\mathbf{k}\eta s} c^{\dagger}_{\mathbf{k}\eta s} [\mathcal{H}_{\mathbf{k}\eta}]_{ss'} c_{\mathbf{k}\eta s'}$, where 
\begin{align}\label{eq:HK}
	\mathcal{H} = \mathcal{H}^{0} + \mathcal{H}^{\text{I}} + \mathcal{H}^{\text{R}} + \mathcal{H}^\text{Z} \, . 
\end{align}
%_{\mathbf{k}\eta}
In Eq.~\eqref{eq:HK}, $\mathcal{H}^{0}$ represents the spin-independent part of the band dispersion, which in the effective mass approximation takes the form 
\begin{align}\label{eq:H_0_warp}
    \mathcal{H}^{0}_{\mathbf{k}\eta} = k^2/2 m + \eta \Delta_\text{W} \left( k/k_\text{F} \right)^3 \cos(3 \varphi)  \,,
\end{align}
It includes in addition to the parabolic dispersion, the trigonal warping $\Delta_\text{W}$, { see Fig.~\ref{fig:model}a}.
Although often \( \Delta _{ \text{W} }  \) is negligible we retain it for a comparison of the SDE in parallel and perpendicular orientations of the magnetic field.
In Eq.~\eqref{eq:H_0_warp} $k_\text{F}$ is the Fermi momentum, at zero SOC and magnetic field, and \( \cos \varphi = \hat{k} \cdot \hat{x} \).

The next two terms of Eq.~\eqref{eq:HK}, $\mathcal{H}^{\ell}$ are the Ising (\( \ell =\text{I} \)) and Rashba (\( \ell =\text{R} \)) SOC, respectively.
We parametrize these terms in a standard way, $\mathcal{H}^{\ell}_{\mathbf{k}\eta} = \boldsymbol{\gamma}^\ell (\mathbf{k} \eta) \cdot  \boldsymbol{\sigma}$, where \( \boldsymbol{\sigma} = ( \sigma_x, \sigma_y, \sigma_z) \) is the vector of Pauli matrices that act in spin space.

The Ising SOC complying to the hexagonal lattice symmetry is shown in Fig.~\ref{fig:model}b.
Close to the $K$ ($K'$) pockets, it simplifies to $\boldsymbol{\gamma}^{\mathrm{I}} (\mathbf{k} \eta)= \eta \Delta_{ \text{I} } \hat{z}$.

When the $m_{\hat{z}}$ symmetry is broken, e.g. by a substrate or by an out-of-plane electric field, the reduction of the $D_{3h}$ symmetry down to $C_{3v}$ allows for Rashba SOC which close to the $K$ ($K'$) pockets takes the form 
\( \boldsymbol{\gamma}^\text{R}(\mathbf{k}\eta) = \alpha_\text{R} \left( \mathbf{k} \times \hat{z} \right) \), see Fig.~\ref{fig:model}c.
Finally, the Zeeman interaction $\mathcal{H}^\text{Z} = \mathbf{B} \cdot \boldsymbol{\sigma}$, where the product of the $g$-factor and Bohr magneton, $\mu_{\mathrm{B}}$ has been absorbed by $\mathbf{B}$.

The local spin singlet pairing interaction, 
\begin{align}\label{eq:H_p}
    V_{\text{p}} = \frac{g}{4} \sum_{\mathbf{q}}\hat{P}_{\mathbf{q}}^\dagger \hat{P}_{\mathbf{q}} ; \,
    \hat{P}_{\mathbf{q}} =  \sum_{\mathbf{k},\eta,s,s'} 
		c_{-\mathbf{k}\bar{\eta} s} \left[ i \sigma_y \right]^\dagger_{s s'} c_{\mathbf{k} + \mathbf{q}\, \eta s'}
\end{align}
is attractive, $g<0$.
 In the $K$ ($K'$)-model the pairing, Eq.~\eqref{eq:H_p} is intervalley as signified by $\bar{\eta} = - \eta$ index.

%\begin{align}\label{eq:H_p}
%    V_{p} = &  \frac{g}{4} \sum_{\mathbf{k},\mathbf{k}',\eta,\eta',s_1,s'_2,s_3,s'_4} 
%		\left\{ c^\dagger_{\mathbf{k} \eta s} \left[ i \sigma_y \right]_{s_1,s_2} c^\dagger_{-\mathbf{k} \bar{\eta} s_2} \right\}
%		\left\{ c_{-\vb{k}'s_3} \left[ i \sigma_y \right]^\dagger_{s_3 s_4} c_{\vb{k}s_4} \right\}\, ,
%\end{align}
%where \( g = \nu_0 \lambda_s \) is the pairing strength and \( \nu_0 = m/2\pi \) is the single electron density of states.

\section{SDE coefficient in GL phenomenology}
\label{sec:SDE}
Here we summarize the way to determine the efficiency of SDE within the GL phenomenology.
To this end, one considers the order parameter \(\Delta \) 
 which is periodic in spatial coordinates, $\mathbf{r}$ with a definite Cooper pair momentum, \(\mathbf{q}\), \(\Delta(\mathbf{r}) = \Delta e^{i \mathbf{q} \mathbf{r}} \).
For a given $\mathbf{q}$ and magnetic field, the GL free energy density reads \cite{Tinkham2004},
\begin{align}\label{eq:fq}
    f(\mathbf{q},\mathbf{B}) = \alpha(\mathbf{q},\mathbf{B}) \Delta^2 + \beta(\mathbf{q},\mathbf{B)}\Delta^4\, .
\end{align}
Finite SDE arises due to the asymmetry, $f(\mathbf{q},\mathbf{B})  \neq f(-\mathbf{q},\mathbf{B})$.
Such asymmetry arises from the odd powers in the expansion of the GL coefficients \cite{edelstein_ginzburg-landau_2021}, 
\begin{align}\label{eq:expand}
		\alpha 
       % \left( \mathbf{q}, \mathbf{B} \right)  
       = 
        \sum_{r=0}^\infty \alpha_r \left( \mathbf{B}, \hat{q} \right)q^r, \, \,\,\,
		\beta 
        % \left( \mathbf{q}, \mathbf{B} \right) 
        = \sum_{r=0}^\infty \beta_r  \left( \mathbf{B}, \hat{q} \right) q^r \, 
\end{align} 
also known as anomalous terms.
In contrast to the even power terms of expansion \eqref{eq:expand}, known as normal, the anomalous terms require breaking of $\mathcal{T}$ and $\mathcal{P}$ symmetries.
Therefore, anomalous contributions are odd in $\mathbf{B}$.
For in-plane $\mathbf{B}$ this implies that $m_{\hat{z}}$ has to be broken for finite anomalous terms.
For an out-of-plane $\mathbf{B}$ the rotation by $\pi$ around $\hat{z}$-axis, $C_{2z}= \mathcal{P} m_{\hat{z}}$ has to be broken while $m_{\hat{z}}$ may be allowed.
In result, for an in-plane $\mathbf{B}$ configuration Rashba SOC is necessary for SDE, while for an out-of-plane field the Ising SOC is sufficient.

Lifshitz invariants, $\alpha_1 (\mathbf{B},\hat{q})q$ in Eq.~\eqref{eq:expand}, are the most studied type of anomalous terms known for some noncetrosymmetric superconductors \cite{Agterberg2012,SmidmanAgterberg17}.
These terms drive the superconductor into the helical state with a finite Cooper pair momentum \( \mathbf{q}_0 \) with  zero current \cite{MineevSamokhin94,MineevSamokhin08}.
In a ring geometry Lifshitz invariant shifts the Little-Parks oscillations \cite{Dimitrova2007}.
Generally, Lifshitz invariant with finite curl produces a large number of observable effects \cite{KapustinRadzihovsky22}.
In particular, in films, Lifshitz invariants produce a thickness-dependent SDE \cite{Kochan23}. 
Here we focus on planar setups such as atomic monolayers, where the Lifshitz invariant alone causes no SDE, because the free energy, \eqref{eq:fq} is symmetric with respect to $\mathbf{q}_0$, \(f(\mathbf{q}_0+ \mathbf{q},\mathbf{B})= f(\mathbf{q}_0-\mathbf{q},\mathbf{B})\).
%Indeed, $\alpha_1 \neq 0$ leaves the free energy, \eqref{eq:fq} symmetric with respect to $\mathbf{q}_0$, \(f(\mathbf{q}_0+ \mathbf{q},\mathbf{B})= f(\mathbf{q}_0-\mathbf{q},\mathbf{B})\).

Within the GL formalism the SDE efficiency reads \cite{Hasan2024}
\begin{align}\label{eq:eta_SDE_GL}
	\eta_{\mathrm{SDE}}\! =\!
		\frac{ 2 \alpha_2 \alpha_3 \beta_0\! -\! 4 \alpha_1 \alpha_4 \beta_0\! -\! \alpha_2^2 \beta_1 \!+ \!\alpha_1 \alpha_2 \beta_2}{ 2 \sqrt{3} \alpha_2^{5/2} \beta_0 } \sqrt{\!-\alpha_0},
\end{align}
where the arguments of \( \alpha, \beta \) are suppressed for brevity. 
Here we compute the expansion coefficients in Eq.~\eqref{eq:eta_SDE_GL} to first order in $B$.

In order to compute the SDE efficiency, we use the standard expressions for the GL coefficients introduced in Eq.~\eqref{eq:fq}, 
\( \alpha = 1/|g| - \Phi_2 /2\), and \( \beta = \Phi_4/4 \) in terms of the correlation functions, 
\begin{align}\label{eq:Phi2m}
    \Phi_{2m} = \sum_{ \mathbf{k},\eta, \varepsilon_n } \mathrm{Tr}\left[
  \sigma_y   G(\mathbf{k}+\mathbf{q} \, \eta,i \varepsilon_n)  \sigma_y  \overline{G}(\mathbf{k}\eta,i \varepsilon_n) \right]^m\, , 
\end{align}
where the Green functions,
\( G \left( \mathbf{k} \eta, \varepsilon_n \right) = \left( i \varepsilon_n - \mathcal{H}_{\mathbf{k}\eta} \right)^{-1} \), \( \overline{ \mathcal{G} }\left( \mathbf{k} \eta, \varepsilon_n \right) = G^{tr} \left( -\mathbf{k}\bar{\eta}, -\varepsilon_n \right) \), and the Matsubara frequency \( \varepsilon_n = 2 \pi T \left( n + 1/2 \right)  \) are standard.

By making an expansion of Eq.~\eqref{eq:Phi2m} in \( q \) up to fourth order and up to first order in $B$ we identify all the coefficients needed to compute the SDE efficiency based on Eq.~\eqref{eq:eta_SDE_GL}.
Similar to \cite{Hasan2024}, we approach the problem in terms of three parameters,
\begin{align}\label{eq:kappas}
    \delta = \alpha_\text{R}/v_\text{F},\,\,\,\,\, \kappa_\text{R} = \alpha_\text{R} k_\text{F} /\pi T_\text{c},\,\,\,\,\, \kappa_\text{I} = \Delta_\text{I}/ \pi T_\text{c},
\end{align}
and evaluate the anomalous GL coefficients to first order in $\delta \ll 1 $.
The other two parameters of the problem satisfy $\kappa_\text{R,I} \gg \delta$, and are otherwise arbitrary.
The latter condition implies that the Fermi energy, $E_\text{F} \gg T_\text{c}$.

%@@@@@@@@@@@@@@@@@@@@@@@@@@@@@@@@@@@@@@@@@@@@@@@@@@@@@@@@@@@@@@@@@@@@@@@@@@@
%@@@@@@@@@@@@@@@@@@@@@@@@@@@@@@@@@@@@@@@@@@@@@@@@@@@@@@@@@@@@@@@@@@@@@@@@@@@
%@@@@@@@@@@@@@@@@@@@@@@@@@@@@@@@@@@@@@@@@@@@@@@@@@@@@@@@@@@@@@@@@@@@@@@@@@@@
\begin{figure}
\begin{center}
\centering
\includegraphics[width=0.45\textwidth]{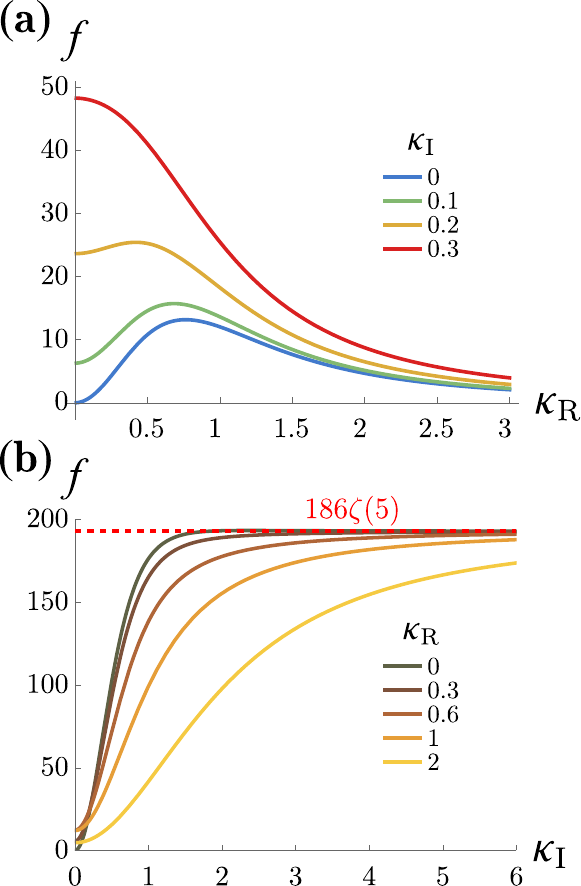}
\caption{
The function \( f \left( \kappa_\text{I},\kappa_\text{R} \right) \) characterizing the SDE efficiency via Eq.~\eqref{eq:eta_SDE}.
(a) The dependence of \( f \) on  \( \kappa_\text{R} \) for a specified values of \( \kappa_\text{I} \). 
(b) The dependence of \( f \) on  \( \kappa_\text{I} \) for a specified values of \( \kappa_\text{R} \). 
\( f \) saturates to a value of 
\( 186 \zeta(5) \approx 192.9  \) at $\kappa_\text{I} \gg \kappa_\text{R}, 1$. 
%This is a total contribution of the intraband contributions in the limit $\kappa_I \gg \kappa_R$.
} \label{fig:function_f}
\end{center}
\end{figure}
%@@@@@@@@@@@@@@@@@@@@@@@@@@@@@@@@@@@@@@@@@@@@@@@@@@@@@@@@@@@@@@@@@@@@@@@@@@@
%@@@@@@@@@@@@@@@@@@@@@@@@@@@@@@@@@@@@@@@@@@@@@@@@@@@@@@@@@@@@@@@@@@@@@@@@@@@
%@@@@@@@@@@@@@@@@@@@@@@@@@@@@@@@@@@@@@@@@@@@@@@@@@@@@@@@@@@@@@@@@@@@@@@@@@@@

%@@@@@@@@@@@@@@@@@@@@@@@@@@@@@@@@@@@@@@@@@@@@@@@@@@@@@@@@@@@@@@@@@@@@@@@@@@@
%@@@@@@@@@@@@@@@@@@@@@@@@@@@@@@@@@@@@@@@@@@@@@@@@@@@@@@@@@@@@@@@@@@@@@@@@@@@
%@@@@@@@@@@@@@@@@@@@@@@@@@@@@@@@@@@@@@@@@@@@@@@@@@@@@@@@@@@@@@@@@@@@@@@@@@@@
\begin{figure}
\begin{center}
\centering
\includegraphics[width=0.46\textwidth]{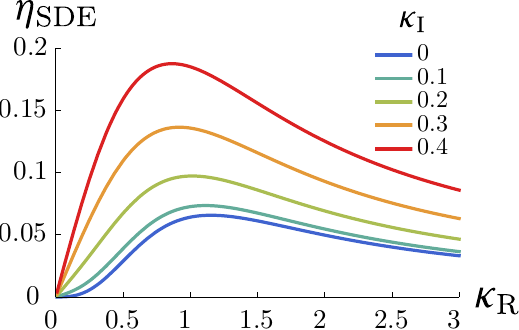}
\caption{
SDE efficiency \( \eta_{\mathrm{SDE}} \) as given by Eq.~\eqref{eq:eta_SDE} divided by \(  \left( \vb{B} \cross \vb{\hat{q}} \right) \cdot \hat{\vb{z}} \sqrt{\frac{T_\text{c}-T}{T_\text{c}}} \frac{T_\text{c}}{ E_\text{F}} \) as a function of \( \kappa_\text{R}\) for different values of \( \kappa_\text{I} \) introduced in Eq.~\eqref{eq:kappas}.
\( \eta_{\mathrm{SDE}} \) vanishes at \( \kappa_\text{R} = 0 \).
At any finite \( \kappa_\text{R} \) the efficiency is strongly enhanced by Ising SOC.
} \label{fig:function_eta}
\end{center}
\end{figure}
%@@@@@@@@@@@@@@@@@@@@@@@@@@@@@@@@@@@@@@@@@@@@@@@@@@@@@@@@@@@@@@@@@@@@@@@@@@@
%@@@@@@@@@@@@@@@@@@@@@@@@@@@@@@@@@@@@@@@@@@@@@@@@@@@@@@@@@@@@@@@@@@@@@@@@@@@
%@@@@@@@@@@@@@@@@@@@@@@@@@@@@@@@@@@@@@@@@@@@@@@@@@@@@@@@@@@@@@@@@@@@@@@@@@@@

\subsection{GL coefficients for in-plane magnetic field }
The normal GL coefficients are even in $B$ according to $\mathcal{T}$ symmetry, and therefore can be evaluated at $B=0$.
They are also not affected by SOC to zero order in $\delta$.
The resulting expressions are standard,
\begin{align}\label{eq:alpha_0}
		\alpha_0 
             = \frac{1}{|g|} -  \pi T \sum_{ \varepsilon_n > 0 } \frac{ 2 \nu_0 }{ \varepsilon_n }\, ,
\end{align}
where $ \nu_0 = m/\pi$ is the density of states per spin includes the contributions of both \(K\) and \( K' \) pockets.
\begin{align}\label{eq:alpha_norm}
\alpha_2 
		& =  \pi T \sum_{ \varepsilon_n > 0 } \frac{ \nu_0 v_\text{F}^2 }{ 4 \varepsilon_n^3 }\, , \,\,
		\alpha_4
			 = - \pi T  \sum_{ \varepsilon_n > 0 } \frac{ 3 \nu_0 v_\text{F}^4 }{ 64 \varepsilon_n^5 } \, ,            
\end{align}
where $v_\text{F} = k_\text{F}/m$ is the Fermi velocity at zero SOC, Zeeman interaction and trigonal warping.
The remaining normal terms entering the expression of the SDE efficiency are
\begin{align}\label{}
		\beta_0 = \pi T \sum_{ \varepsilon_n > 0 } \frac{ \nu_0 }{ 2 \varepsilon_n^3 }\, , \,\,\,
		\beta_2  = -  \pi T \sum_{ \varepsilon_n > 0 } \frac{ 3 \nu_0 v_\text{F}^2 }{ 8 \varepsilon_n^5 } \, .
\end{align}

We now turn to anomalous coefficients. The Lifshitz invariant
\begin{align}\label{}
	\alpha_1 
		= \left( \vb{B} \cross \vu{q} \right) \cdot \vu{z} \alpha_\text{R}  \left(\frac{ 2 \kappa_\text{I}^2 + \kappa_\text{R}^2 }{ \kappa_\text{I}^2 +\kappa_\text{R}^2 }\right) (\alpha_1^{\text{intra}} + \alpha_1^{\text{inter}})
\end{align}
naturally splits into two distinct contributions, 
\begin{subequations}\label{eq:alpha1}
\begin{align}\label{eq:alpha1a}
	\alpha_1^\text{intra}
		= & - \pi T \sum_{ \varepsilon_n > 0 } \frac{ \nu_0 }{ \varepsilon_n^3 }\, ,
	\end{align}
    \begin{align}\label{eq:alpha1b}
	\alpha_1^\text{inter}
		= &  \pi T \sum_{ \varepsilon_n > 0 } \frac{ \nu_0 }{ \varepsilon_n \left( \pi^2 T^2 \kappa^2 + \varepsilon_n^2 \right) } \, ,
\end{align}
\end{subequations}
where $\kappa = \sqrt{\kappa_\text{I}^2 + \kappa_\text{R}^2}$.
The intraband contribution, Eq.~\eqref{eq:alpha1a} results from the pairing of particles on the outer (inner) Fermi pockets.
 The second, interband contribution Eq.~\eqref{eq:alpha1b} arises from pairing of particles at inner and outer Fermi surfaces, see Fig.~\ref{fig:model}a. 
Unlike the intraband contributions, the interband contribution is suppressed in the limit of strong SOC, $\kappa \gg 1$.
Note that the two contributions cancel each other out in the limit $\kappa=0$.
This cancellation is not accidental and follows from the gauge invariance as discussed in Sec.~\ref{sec:Intro}.

Similarly to the Lifshitz invariant, Eq.~\eqref{eq:alpha1}, we present the third order anomalous term as a sum of the intra- and interband contributions, 
\begin{align}
    \alpha_3
         = \left( \vb{B} \cross \vu{q} \right)\! \cdot\! \vu{z} \alpha_\text{R} \left(\frac{4 \kappa_\text{I}^2 + \kappa_\text{R}^2 }{ \kappa_\text{I}^2 +\kappa_\text{R}^2 }\right) \left( \alpha_3^\text{intra}+\alpha_3^\text{inter} \right) ,
\end{align}
\begin{subequations}
\begin{align}
    \alpha_3^\text{intra} 
        = \pi T \sum_{ \varepsilon_n > 0 } \frac{ 3 \nu_0 v_\text{F}^2 }{ 8 \varepsilon_n^5 }\, ,
\end{align}
\begin{align}
    \alpha_3^\text{inter}
       = &  - \pi T \sum_{ \varepsilon_n > 0 } \frac{ \nu_0 v_\text{F}^2 }{ 16 
        \varepsilon_n^3 \left[ \pi^2 T^2 \kappa^2 + \varepsilon_n^2 \right]^3 }
        \notag\\
        &\times 
        \left[ 6 \varepsilon_n^4 + 3 \varepsilon_n^2 \pi^2 T^2 \kappa^2  + \pi^4 T^4 \kappa^4 \right]\, .
\end{align}
\end{subequations}

The remaining anomalous coefficient \( \beta_1 \) reads, 
\begin{align}\label{}
	\beta_1 
		= \left( \vb{B} \cross \vu{q} \right) \cdot \vu{z} \alpha_\text{R} \left(\frac{ 2 \kappa_\text{I}^2 + \kappa_\text{R}^2 }{ \kappa_\text{I}^2 +\kappa_\text{R}^2 }\right) \left( \beta_3^\text{intra} + \beta_3^\text{inter} \right), 
\end{align}
\begin{subequations}
\begin{align}\label{}
	\beta_1^\text{intra}
		=  \pi T \sum_{ \varepsilon_n > 0 } \frac{ 3 \nu_0 }{ 2 \varepsilon_n^5 }\, ,
        \end{align}
\begin{align}
	\beta_1^\text{inter}
		= - \pi T \sum_{ \varepsilon_n > 0 } \frac{ \nu_0 \left[ \pi^2 T^2\kappa^2 + 3 \varepsilon_n^2 \right]
        }{ 2 \varepsilon_n^3 \left[ \pi^2 T^2 \kappa^2 + \varepsilon_n^2 \right]^2 } \,.
\end{align}
\end{subequations}
Similar to the Lifshitz invariant, the intra- and interband contributions to $\alpha_3$ and $\beta_1$ cancel each other out for $\kappa=0$.
%%%%%%%%%%%%%%%%%%%%%%%%%%%%
\subsection{SDE efficiency}
\label{sec:SDE_efficiency}
Having calculated all the necessary GL coefficients, we use Eq.~\eqref{eq:eta_SDE_GL} to write the SDE efficiency as 
\begin{align}\label{eq:eta_SDE}
    \eta_{\mathrm{SDE}}
        = \frac{ \left( \vb{B} \cross \vb{\hat{q}} \right) \cdot \hat{\vb{z}} \delta }{ 7 \pi T_\text{c} \sqrt{42\zeta(3)^3} } \sqrt{\frac{T_\text{c}-T}{T_\text{c}}}
         f \left( \kappa_\text{I},\kappa_\text{R} \right) \, .
\end{align}
The function \(   f \left( \kappa_\text{I},\kappa_\text{R} \right)   \) is a measure of the SDE efficience.
We present it for different fixed values of \( \kappa_\text{I} \) ($\kappa_\text{R}$) as a function of \( \kappa_\text{R} \) (\( \kappa_\text{I} \)) in Fig.~\ref{fig:function_f}, panel a (b).
The enhancement of the SDE efficiency by Ising SOC is shown directly in Fig.~\ref{fig:function_eta}.

In the same way as with the GL coefficients, the function $f \left( \kappa_\text{I},\kappa_\text{R} \right)$ splits into intra- and interband parts, \( f =  f^\text{intra} + f^\text{inter} \).
The intraband contribution takes the form,
\begin{subequations}\label{eq:f}
    \begin{align}\label{eq:fintra}
        f^\text{intra}
            = & 186 \zeta(5) \kappa_\text{I}^2 / \left( \kappa_\text{I}^2 + \kappa_\text{R}^2 \right) \,,
    \end{align}
and the interband part reads,
\begin{align}\label{eq:finter}
    f^\text{inter}
        = & - \frac{ 4 \kappa_\text{I}^2 + \kappa_\text{R}^2 }{ \kappa^4 } \Re{ \psi^{(2)} \left( \frac{1}{2} - \frac{i \kappa}{2} \right) } + 14 \frac{ \zeta(3)}{ \kappa^4}  \kappa_\text{R}^2
        \notag \\
        & + 16 \frac{ 2 \kappa_\text{I}^2 + \kappa_\text{R}^2 }{ \kappa^6 } \Re{ \psi^{(0)} \left( \frac{1}{2} - \frac{i\kappa}{2}  \right) - \psi^{(0)} \left( \frac{1}{2}\right) } 
        \notag \\
        & + 8 \frac{ 2 \kappa_\text{I}^2 + \kappa_\text{R}^2}{ \kappa^5 } \Re{ i\psi^{(1)} \left( \frac{1}{2} - \frac{i\kappa}{2} \right) } 
     \,,
\end{align}
\end{subequations}
with \( \psi^{(n)} \) being the n-th polygamma function. 

\begin{figure}
\begin{center}
\centering
\includegraphics[width=0.45\textwidth]{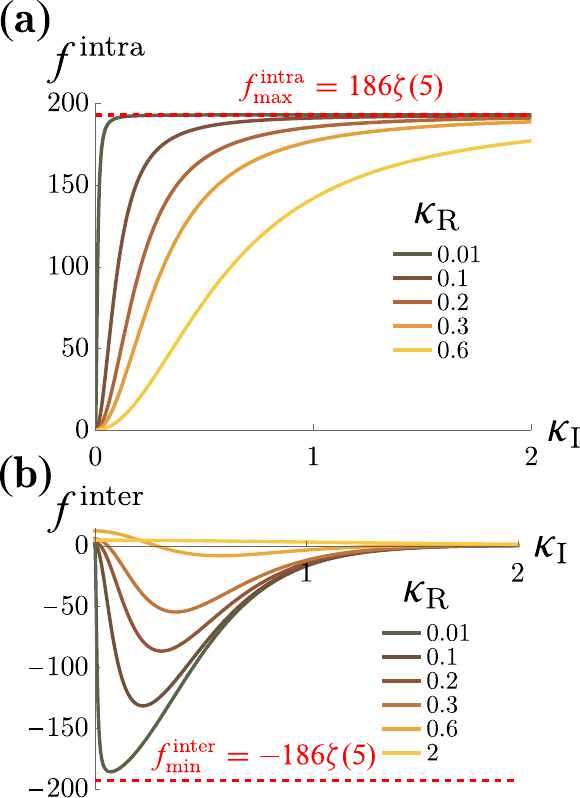}
\caption{
Intraband ($f^\text{intra}$) and interband ($f^\text{inter}$) contributions to the function \( f \) [see Eq.~\eqref{eq:f}] that determines the SDE efficiency, via Eq.~\eqref{eq:eta_SDE},  as a function of \( \kappa_\text{I} \) at different values of \( \kappa_\text{R} \). 
(a) The intraband saturates when the Ising SOC dominates the Rashba SOC \( \kappa_\text{I} \gg \kappa_\text{R} \).
(b) 
The interband contribution is suppressed at $\kappa \gg 1$. It roughly cancels the intraband contribution in the opposite limit, $\kappa \ll 1$.
As Rashba SOC gets weaker, larger values of \( \kappa_\text{I} \) are consistent with this condition. 
This implies that as \( \kappa_\text{R} \) decreases the ratio \( \kappa_\text{I}/\kappa_\text{R} \) consistent with $\kappa \ll 1$ can exceed one.
Since intraband contribution approaches $f_\text{max}^\text{intra}$ at \( \kappa_\text{I}/\kappa_\text{R} \gg 1 \) the interband contribution reaches the minimum value \( f_\text{min}^\text{inter} = -f_\text{max}^\text{intra} \) at decreasing values of \( \kappa_\text{I} \) as \( \kappa_\text{R} \) decreases.
} 
\label{fig:intra_inter}
\end{center}
\end{figure}

In a Rashba superconductor, \(\kappa_\text{I} =0 \), intraband contribution, Eq.~\eqref{eq:fintra} vanishes, \(f^\text{intra} \left( \kappa_\text{I} = 0,\kappa_\text{R} \right)=0 \).
We find that it is finite for any $\kappa_\text{I} \neq 0$.
As shown in Fig.~\ref{fig:intra_inter}a these contributions saturates to a universal value, $f_\text{max}^\text{intra}= 186 \zeta(5)\approx 193$ for $\kappa_\text{I} \gg \kappa_\text{R}$. 
According to Eq.~\eqref{eq:fintra} the saturation at $90 \%$ is reached at $\kappa_\text{I} = 3 \kappa_\text{R}$.

%The interband contributions are generally suppressed at large SOC, $\kappa \gg 1$. 
At large SOC, $\kappa \gtrsim  1$  the interband contribution is suppressed.
In the opposite limit, $\kappa \lesssim  1$ the interband contribution is not negligible.
It tends to cancel the intraband contribution at small $\kappa$.
This cancellation is systematic and holds for each of the anomalous coefficients, $\alpha_1$, $\alpha_3$ and $\beta_1$ separately; see Sect.~\ref{sec:Intro} for more details.
Let us define $\kappa_*$ such that for $\kappa < \kappa_*$ the cancellation holds within less than $10 \%$.

To estimate $\kappa_*$ we focus on the intersection of curves with the vertical axis (\( \kappa_\text{R}=0\)) in Fig.~\ref{fig:function_f}a.
At \( \kappa_\text{R}=0\), $\kappa = \kappa_\text{I}$, and the $\kappa$-independent intraband contribution is at maximum of $f_\text{max}^\text{intra} = 186 \zeta(5)$.
Observe that at $\kappa_\text{I}=0.2$, $f$ reaches approximately 10$\%$ of $f_\text{max}^\text{intra}$.
We therefore, estimate $\kappa_* = 0.2$.

In summary, the interband contribution exceeds $90 \%$ of $f_\text{max}^\text{intra}$ in magnitude, provided that the two inequalities
$\kappa_\text{I}^2 + \kappa_\text{R}^2 < \kappa_*^2$ and $\kappa_\text{I} > 3 \kappa_\text{R}$ are satisfied. 
For $\kappa_\text{R} < \kappa_* / \sqrt{10} \approx 0.06$ it is possible for some \( \kappa_\text{I} \).
We illustrate these features in Fig.~\ref{fig:intra_inter}b.

%@@@@@@@@@@@@@@@@@@@@@@@@@@@@@@@@@@@@@@@@@@@@@@@@@@@@@@@@@@@@@@@@@@@@@@@@@@@
%@@@@@@@@@@@@@@@@@@@@@@@@@@@@@@@@@@@@@@@@@@@@@@@@@@@@@@@@@@@@@@@@@@@@@@@@@@@
%@@@@@@@@@@@@@@@@@@@@@@@@@@@@@@@@@@@@@@@@@@@@@@@@@@@@@@@@@@@@@@@@@@@@@@@@@@@

%@@@@@@@@@@@@@@@@@@@@@@@@@@@@@@@@@@@@@@@@@@@@@@@@@@@@@@@@@@@@@@@@@@@@@@@@@@@
%@@@@@@@@@@@@@@@@@@@@@@@@@@@@@@@@@@@@@@@@@@@@@@@@@@@@@@@@@@@@@@@@@@@@@@@@@@@
%@@@@@@@@@@@@@@@@@@@@@@@@@@@@@@@@@@@@@@@@@@@@@@@@@@@@@@@@@@@@@@@@@@@@@@@@@@@
\section{Discussion}
\label{sec:Discussion}
We have proposed a mechanism for the SDE in Ising superconductors, specifically for an in-plane magnetic field configuration. 
This mechanism is distinct from the SDE driven by a perpendicular field, \( \mathbf{B}_\perp \parallel \hat{z} \), recently observed in few-layer NbSe$_2$ \cite{BauriedlParadiso22}. 
In that realization, the superconducting NbSe$_2$ is symmetrically encapsulated by h-BN from both the top and bottom, preserving the \( m_{\hat{z}} \) symmetry and preventing Rashba SOC. 
The observed finite \( \eta^\perp_{\mathrm{SDE}} \) in this experiment is a manifestation of normal-state magnetochiral anisotropy \cite{Rikken:PRL2001,WakatsukiNagaosa17,WakatsukiNagaosa18,LeggLossKlinovaja22}.

The crossover from Ising-dominated (\(\kappa_\text{I} > \kappa_\text{R}\)) to Rashba-dominated (\(\kappa_\text{R} > \kappa_\text{I}\)) regimes has been reported in epitaxial Bi$_2$Se$_3$/monolayer NbSe$_2$ heterostructures \cite{Yi2022}. 
This system has been reported to be superconducting and could serve as a natural platform for realizing an enhanced SDE.

NbSe$_2$ monolayer on a hexagonal boron-nitride layer on a SiO$_2$/Si substrate exhibits a six-fold angular variation of the in-plane critical field \cite{Cho2022}.  
When the \(m_{\hat{z}}\) symmetry is intact, the system is isotropic with respect to the orientation of the in-plane field \cite{Attias2024}.  
Therefore, this provides indirect evidence of the Rashba SOC in this heterostructure \cite{Haim2022}.

Rashba SOC can be introduced in an Ising superconductor monolayer by gating as well.  
This produces an out-of-plane electric field, breaking the \(m_{\hat{z}}\) symmetry.  
This has been achieved in MoSe$_2$ and MoS$_2$ monolayers \cite{Cheng_2018}.

To estimate a minimal Rashba SOC required for obtaining a noticeable SDE we make comparison of the measured SDE in a perpendicular field reported in Ref.~\cite{BauriedlParadiso22}.
In this case, the SDE due to the spin degrees of freedom follows from the trigonal warping term \(\propto \Delta_\text{W}\) in Eq.~\eqref{eq:H_0_warp}.
The GL formalism is easily adapted to this scenario.
The only anomalous term allowed by \(D_{3h}\) symmetry, 
\(\alpha_3^\perp \propto B_\perp(q_x^3 - 3 q_x q_y^2)/q^3\) has been computed in Ref.~\cite{WakatsukiNagaosa17}.
This gives, using Eq.~\eqref{eq:eta_SDE_GL},
\begin{equation}\label{eq:eta_perp}
   \eta^\perp_{\mathrm{SDE}} = \frac{ 124 \sqrt{ 6/7 } \zeta(5) }{ 28 \sqrt{ \zeta(3)^3} \pi }\frac{\Delta_\text{I} \Delta_\text{W}}{T_\text{c} E_\text{F}^2}   \! B_\perp \!\frac{\left( q_x^3 - 3 q_x q_y^2 \right)}{q^3} \sqrt{ \frac{T_\text{c}-T}{T_\text{c}} },
\end{equation}
where the numerical prefactor \( \approx 1.03 \).
Next we compare the efficiency, Eq.~\eqref{eq:eta_perp}, to the efficiency in the parallel field, $\mathbf{B}_{\parallel} \perp \hat{z}$ summarized in Sec.~\ref{sec:SDE_efficiency}.
For definiteness, we consider $\mathbf{q} \parallel \hat{x}$ and $\mathbf{B}_{\parallel} \parallel \hat{y}$.
Normally, Ising SOC is strong, \( \kappa_\text{I} \gg 1 \), and the interband contribution, Eq.~\eqref{eq:finter} is negligible: $f^\text{intra} \gg f^{\text{inter}}$.
As the strong Ising SOC dominates the Rashba SOC, \( \kappa_\text{I} \gg \kappa_\text{R} \), the SDE efficiency is given by Eq.~\eqref{eq:eta_SDE} with $f(\kappa_\text{I},\kappa_\text{R})\approx 186 \zeta(5)$.
The ratio of the SDE efficiencies at parallel and perpendicular field orientations, 
\begin{align}\label{eq:SDE_ratio}
    \frac{\eta_\text{SDE}}{\eta^\perp_\text{SDE}}
        = \frac{ 1 }{ 2 }\frac{ B_\parallel }{ B_\perp }
         \frac{ E_\text{F}  }{ \Delta_\text{I}  }
         \frac{\Delta_\text{R}}{\Delta_\text{W}}\,,
\end{align}
is fixed by the ratio of Rashba spin splitting and trigonal warping; and the ratio of the Fermi energy and the Ising spin splitting.
A characteristic Ising SOC for TMDs is \( \Delta_\text{I} \approx 0.05-\qty{0.12}{\electronvolt} \) \cite{delaBarrera2018}, and their Fermi level exceeded \( E_\text{F} > \qty{1}{\electronvolt} \) \cite{Xi2015,Zhang2014}. Thus we conservatively estimate the first ratio \( E_\text{F} / \Delta_\text{I} \gtrsim 10 \).

Based on Density Functional Theory, Ref.~\cite{Kormanyos_2015} reports $C_{3w} =  \Delta_\text{W}/k_\text{F}^3 \approx \qty{5}{ \electronvolt \angstrom^{3}}$ for the valence band of MoS$_2$ and MoSe$_2$. 
The wavevector, \(\mathbf{k}_K\) at \( K \) point satisfies \( \abs{\vb{k}_K} = 4 \pi a^{-1} /3 \approx \qty{1.3}{\angstrom^{-1}} \) for a characteristic lattice constant of \( a = \qty{3.2}{\angstrom} \).
Since our model holds for small Fermi surfaces, we set \( k_\text{F} = 0.1 \abs{\vb{k}_\text{K}} \), and estimate \( \Delta_\text{W} \approx \qty{11}{m\electronvolt} \).

Therefore, for the parallel field efficiency to exceed the perpendicular efficiency in Eq.~\eqref{eq:SDE_ratio}, \( \Delta_\text{R} \gtrsim \qty{2}{\milli \electronvolt} \)  is required. 
According to Ref.~\cite{Cheng_2018} the Rashba splitting  greatly exceeds the above requirement in gated MoS$_2$ and MoSe$_2$.
Ion gating is known to induce superconductivity in MoS$_2$ \cite{lu2015evidence,costanzo2018tunnelling,saito2016superconductivity}.
This shows that our setup is a realistic way to achieve a large SDE in parallel magnetic field, e.g. in gated MoS$_2$.

\section{Summary}
\label{sec:Conclusions}
In summary, we have proposed the two-dimensional Ising superconductor as a platform for the SDE with large efficiency driven by an in-plane magnetic field.
Similar to the Rashba superconductor, finite Rashba spin splitting is necessary for finite SDE.
Such splitting is achieved experimentally by growing the Ising superconductor on a substrate or by gating.

With minimal Rashba splitting, the Ising superconductor produces an SDE with efficiency that greatly exceeds the maximal efficiency of the Rashba superconductor.
Moreover, realistically it can be comparable or greater than the efficiency in a perpendicular magnetic field.
 
In this work, we focus on homogeneous systems.
It is of interest to implore the possibility of the SDE in the Josephson junction geometry. 
The in-plane exchange field has been suggested to strongly affect junctions with an Ising superconductor sandwiched by two ferromagnetic electrodes \cite{Tang2021}.
In this case, the properties of the junction are affected by the triplet correlation induced by the in-plane field into an otherwise singlet superconductor \cite{Mockli2019,Tang2021a}. 
It is of interest to explore these research directions in the future.

\section*{Acknowledgements}

We thank Y. Anahory, B. Andersen, M. Haim, J. Hasan, A. Osin, D. Shaffer, H. Shteinberg and M. Tsindlekht for useful discussions. 
L. A., I. B., and M. K. acknowledge the financial support from the Israel Science Foundation, Grant No. 2665/20. The work of A. L. was supported by the National Science Foundation Grant No. DMR-2452658 and H. I. Romnes Faculty Fellowship provided by the University of Wisconsin-Madison Office of the Vice Chancellor for Research and Graduate Education with funding from the Wisconsin Alumni Research Foundation.

%\bibliography{biblio}

\begin{thebibliography}{69}%
\makeatletter
\providecommand \@ifxundefined [1]{%
 \@ifx{#1\undefined}
}%
\providecommand \@ifnum [1]{%
 \ifnum #1\expandafter \@firstoftwo
 \else \expandafter \@secondoftwo
 \fi
}%
\providecommand \@ifx [1]{%
 \ifx #1\expandafter \@firstoftwo
 \else \expandafter \@secondoftwo
 \fi
}%
\providecommand \natexlab [1]{#1}%
\providecommand \enquote  [1]{``#1''}%
\providecommand \bibnamefont  [1]{#1}%
\providecommand \bibfnamefont [1]{#1}%
\providecommand \citenamefont [1]{#1}%
\providecommand \href@noop [0]{\@secondoftwo}%
\providecommand \href [0]{\begingroup \@sanitize@url \@href}%
\providecommand \@href[1]{\@@startlink{#1}\@@href}%
\providecommand \@@href[1]{\endgroup#1\@@endlink}%
\providecommand \@sanitize@url [0]{\catcode `\\12\catcode `\$12\catcode `\&12\catcode `\#12\catcode `\^12\catcode `\_12\catcode `\%12\relax}%
\providecommand \@@startlink[1]{}%
\providecommand \@@endlink[0]{}%
\providecommand \url  [0]{\begingroup\@sanitize@url \@url }%
\providecommand \@url [1]{\endgroup\@href {#1}{\urlprefix }}%
\providecommand \urlprefix  [0]{URL }%
\providecommand \Eprint [0]{\href }%
\providecommand \doibase [0]{https://doi.org/}%
\providecommand \selectlanguage [0]{\@gobble}%
\providecommand \bibinfo  [0]{\@secondoftwo}%
\providecommand \bibfield  [0]{\@secondoftwo}%
\providecommand \translation [1]{[#1]}%
\providecommand \BibitemOpen [0]{}%
\providecommand \bibitemStop [0]{}%
\providecommand \bibitemNoStop [0]{.\EOS\space}%
\providecommand \EOS [0]{\spacefactor3000\relax}%
\providecommand \BibitemShut  [1]{\csname bibitem#1\endcsname}%
\let\auto@bib@innerbib\@empty
%</preamble>
\bibitem [{\citenamefont {Jiang}\ and\ \citenamefont {Hu}(2022)}]{jiang_superconducting_2022}%
  \BibitemOpen
  \bibfield  {author} {\bibinfo {author} {\bibfnamefont {K.}~\bibnamefont {Jiang}}\ and\ \bibinfo {author} {\bibfnamefont {J.}~\bibnamefont {Hu}},\ }\bibfield  {title} {\bibinfo {title} {Superconducting diode effects},\ }\href@noop {} {\bibfield  {journal} {\bibinfo  {journal} {Nature Physics}\ ,\ \bibinfo {pages} {1}} (\bibinfo {year} {2022})},\ \bibinfo {note} {publisher: Nature Publishing Group}\BibitemShut {NoStop}%
\bibitem [{\citenamefont {Nadeem}\ \emph {et~al.}(2023)\citenamefont {Nadeem}, \citenamefont {Fuhrer},\ and\ \citenamefont {Wang}}]{Nadeem2023}%
  \BibitemOpen
  \bibfield  {author} {\bibinfo {author} {\bibfnamefont {M.}~\bibnamefont {Nadeem}}, \bibinfo {author} {\bibfnamefont {M.~S.}\ \bibnamefont {Fuhrer}},\ and\ \bibinfo {author} {\bibfnamefont {X.}~\bibnamefont {Wang}},\ }\bibfield  {title} {\bibinfo {title} {The superconducting diode effect},\ }\href@noop {} {\bibfield  {journal} {\bibinfo  {journal} {Nature Reviews Physics}\ }\textbf {\bibinfo {volume} {5}},\ \bibinfo {pages} {558} (\bibinfo {year} {2023})}\BibitemShut {NoStop}%
\bibitem [{\citenamefont {Zinkl}\ \emph {et~al.}(2022)\citenamefont {Zinkl}, \citenamefont {Hamamoto},\ and\ \citenamefont {Sigrist}}]{ZinklHamamotoSigrist22}%
  \BibitemOpen
  \bibfield  {author} {\bibinfo {author} {\bibfnamefont {B.}~\bibnamefont {Zinkl}}, \bibinfo {author} {\bibfnamefont {K.}~\bibnamefont {Hamamoto}},\ and\ \bibinfo {author} {\bibfnamefont {M.}~\bibnamefont {Sigrist}},\ }\bibfield  {title} {\bibinfo {title} {Symmetry conditions for the superconducting diode effect in chiral superconductors},\ }\href@noop {} {\bibfield  {journal} {\bibinfo  {journal} {Physical Review Research}\ }\textbf {\bibinfo {volume} {4}},\ \bibinfo {pages} {033167} (\bibinfo {year} {2022})},\ \bibinfo {note} {publisher: American Physical Society}\BibitemShut {NoStop}%
\bibitem [{\citenamefont {Levitov}\ \emph {et~al.}(1985)\citenamefont {Levitov}, \citenamefont {Nazarov},\ and\ \citenamefont {Eliashberg}}]{Levitov1985}%
  \BibitemOpen
  \bibfield  {author} {\bibinfo {author} {\bibfnamefont {L.~S.}\ \bibnamefont {Levitov}}, \bibinfo {author} {\bibfnamefont {Y.~V.}\ \bibnamefont {Nazarov}},\ and\ \bibinfo {author} {\bibfnamefont {G.~M.}\ \bibnamefont {Eliashberg}},\ }\bibfield  {title} {\bibinfo {title} {Magnetostatics of superconductors without an inversion center},\ }\href@noop {} {\bibfield  {journal} {\bibinfo  {journal} {JETP Letters}\ }\textbf {\bibinfo {volume} {41}},\ \bibinfo {pages} {445} (\bibinfo {year} {1985})}\BibitemShut {NoStop}%
\bibitem [{\citenamefont {Edelstein}(1996)}]{Edelstein1996}%
  \BibitemOpen
  \bibfield  {author} {\bibinfo {author} {\bibfnamefont {V.~M.}\ \bibnamefont {Edelstein}},\ }\bibfield  {title} {\bibinfo {title} {The {G}inzburg-{L}andau equation for superconductors of polar symmetry},\ }\href@noop {} {\bibfield  {journal} {\bibinfo  {journal} {Journal of Physics: Condensed Matter}\ }\textbf {\bibinfo {volume} {8}},\ \bibinfo {pages} {339} (\bibinfo {year} {1996})}\BibitemShut {NoStop}%
\bibitem [{\citenamefont {Ando}\ \emph {et~al.}(2020)\citenamefont {Ando}, \citenamefont {Miyasaka}, \citenamefont {Li}, \citenamefont {Ishizuka}, \citenamefont {Arakawa}, \citenamefont {Shiota}, \citenamefont {Moriyama}, \citenamefont {Yanase},\ and\ \citenamefont {Ono}}]{Ando2020}%
  \BibitemOpen
  \bibfield  {author} {\bibinfo {author} {\bibfnamefont {F.}~\bibnamefont {Ando}}, \bibinfo {author} {\bibfnamefont {Y.}~\bibnamefont {Miyasaka}}, \bibinfo {author} {\bibfnamefont {T.}~\bibnamefont {Li}}, \bibinfo {author} {\bibfnamefont {J.}~\bibnamefont {Ishizuka}}, \bibinfo {author} {\bibfnamefont {T.}~\bibnamefont {Arakawa}}, \bibinfo {author} {\bibfnamefont {Y.}~\bibnamefont {Shiota}}, \bibinfo {author} {\bibfnamefont {T.}~\bibnamefont {Moriyama}}, \bibinfo {author} {\bibfnamefont {Y.}~\bibnamefont {Yanase}},\ and\ \bibinfo {author} {\bibfnamefont {T.}~\bibnamefont {Ono}},\ }\bibfield  {title} {\bibinfo {title} {Observation of superconducting diode effect},\ }\href@noop {} {\bibfield  {journal} {\bibinfo  {journal} {Nature}\ }\textbf {\bibinfo {volume} {584}},\ \bibinfo {pages} {373} (\bibinfo {year} {2020})},\ \bibinfo {note} {number: 7821 Publisher: Nature Publishing Group}\BibitemShut {NoStop}%
\bibitem [{\citenamefont {Baumgartner}\ \emph {et~al.}(2022)\citenamefont {Baumgartner}, \citenamefont {Fuchs}, \citenamefont {Costa}, \citenamefont {Reinhardt}, \citenamefont {Gronin}, \citenamefont {Gardner}, \citenamefont {Lindemann}, \citenamefont {Manfra}, \citenamefont {Faria~Junior}, \citenamefont {Kochan}, \citenamefont {Fabian}, \citenamefont {Paradiso},\ and\ \citenamefont {Strunk}}]{BaumgartnerManfraStrunk22}%
  \BibitemOpen
  \bibfield  {author} {\bibinfo {author} {\bibfnamefont {C.}~\bibnamefont {Baumgartner}}, \bibinfo {author} {\bibfnamefont {L.}~\bibnamefont {Fuchs}}, \bibinfo {author} {\bibfnamefont {A.}~\bibnamefont {Costa}}, \bibinfo {author} {\bibfnamefont {S.}~\bibnamefont {Reinhardt}}, \bibinfo {author} {\bibfnamefont {S.}~\bibnamefont {Gronin}}, \bibinfo {author} {\bibfnamefont {G.~C.}\ \bibnamefont {Gardner}}, \bibinfo {author} {\bibfnamefont {T.}~\bibnamefont {Lindemann}}, \bibinfo {author} {\bibfnamefont {M.~J.}\ \bibnamefont {Manfra}}, \bibinfo {author} {\bibfnamefont {P.~E.}\ \bibnamefont {Faria~Junior}}, \bibinfo {author} {\bibfnamefont {D.}~\bibnamefont {Kochan}}, \bibinfo {author} {\bibfnamefont {J.}~\bibnamefont {Fabian}}, \bibinfo {author} {\bibfnamefont {N.}~\bibnamefont {Paradiso}},\ and\ \bibinfo {author} {\bibfnamefont {C.}~\bibnamefont {Strunk}},\ }\bibfield  {title} {\bibinfo {title} {Supercurrent rectification and magnetochiral effects in symmetric {Josephson} junctions},\ }\href@noop {} {\bibfield
  {journal} {\bibinfo  {journal} {Nature Nanotechnology}\ }\textbf {\bibinfo {volume} {17}},\ \bibinfo {pages} {39} (\bibinfo {year} {2022})}\BibitemShut {NoStop}%
\bibitem [{\citenamefont {Lotfizadeh}\ \emph {et~al.}(2024)\citenamefont {Lotfizadeh}, \citenamefont {Schiela}, \citenamefont {Pekerten}, \citenamefont {Yu}, \citenamefont {Elfeky}, \citenamefont {Strickland}, \citenamefont {Matos-Abiague},\ and\ \citenamefont {Shabani}}]{LotfizadehShabani24}%
  \BibitemOpen
  \bibfield  {author} {\bibinfo {author} {\bibfnamefont {N.}~\bibnamefont {Lotfizadeh}}, \bibinfo {author} {\bibfnamefont {W.~F.}\ \bibnamefont {Schiela}}, \bibinfo {author} {\bibfnamefont {B.}~\bibnamefont {Pekerten}}, \bibinfo {author} {\bibfnamefont {P.}~\bibnamefont {Yu}}, \bibinfo {author} {\bibfnamefont {B.~H.}\ \bibnamefont {Elfeky}}, \bibinfo {author} {\bibfnamefont {W.~M.}\ \bibnamefont {Strickland}}, \bibinfo {author} {\bibfnamefont {A.}~\bibnamefont {Matos-Abiague}},\ and\ \bibinfo {author} {\bibfnamefont {J.}~\bibnamefont {Shabani}},\ }\bibfield  {title} {\bibinfo {title} {Superconducting diode effect sign change in epitaxial {Al}-{InAs} {Josephson} junctions},\ }\href {https://doi.org/10.1038/s42005-024-01618-5} {\bibfield  {journal} {\bibinfo  {journal} {Communications Physics}\ }\textbf {\bibinfo {volume} {7}},\ \bibinfo {pages} {1} (\bibinfo {year} {2024})},\ \bibinfo {note} {publisher: Nature Publishing Group}\BibitemShut {NoStop}%
\bibitem [{\citenamefont {Yerin}\ \emph {et~al.}(2024)\citenamefont {Yerin}, \citenamefont {Drechsler}, \citenamefont {Varlamov}, \citenamefont {Cuoco},\ and\ \citenamefont {Giazotto}}]{Erin2024}%
  \BibitemOpen
  \bibfield  {author} {\bibinfo {author} {\bibfnamefont {Y.}~\bibnamefont {Yerin}}, \bibinfo {author} {\bibfnamefont {S.-L.}\ \bibnamefont {Drechsler}}, \bibinfo {author} {\bibfnamefont {A.~A.}\ \bibnamefont {Varlamov}}, \bibinfo {author} {\bibfnamefont {M.}~\bibnamefont {Cuoco}},\ and\ \bibinfo {author} {\bibfnamefont {F.}~\bibnamefont {Giazotto}},\ }\bibfield  {title} {\bibinfo {title} {Supercurrent rectification with time-reversal symmetry broken multiband superconductors},\ }\href {https://doi.org/10.1103/PhysRevB.110.054501} {\bibfield  {journal} {\bibinfo  {journal} {Phys. Rev. B}\ }\textbf {\bibinfo {volume} {110}},\ \bibinfo {pages} {054501} (\bibinfo {year} {2024})}\BibitemShut {NoStop}%
\bibitem [{\citenamefont {Debnath}\ and\ \citenamefont {Dutta}(2024)}]{Debnath2024}%
  \BibitemOpen
  \bibfield  {author} {\bibinfo {author} {\bibfnamefont {D.}~\bibnamefont {Debnath}}\ and\ \bibinfo {author} {\bibfnamefont {P.}~\bibnamefont {Dutta}},\ }\bibfield  {title} {\bibinfo {title} {Gate-tunable josephson diode effect in rashba spin-orbit coupled quantum dot junctions},\ }\href {https://doi.org/10.1103/PhysRevB.109.174511} {\bibfield  {journal} {\bibinfo  {journal} {Phys. Rev. B}\ }\textbf {\bibinfo {volume} {109}},\ \bibinfo {pages} {174511} (\bibinfo {year} {2024})}\BibitemShut {NoStop}%
\bibitem [{\citenamefont {Miyasaka}\ \emph {et~al.}(2021)\citenamefont {Miyasaka}, \citenamefont {Kawarazaki}, \citenamefont {Narita}, \citenamefont {Ando}, \citenamefont {Ikeda}, \citenamefont {Hisatomi}, \citenamefont {Daido}, \citenamefont {Shiota}, \citenamefont {Moriyama}, \citenamefont {Yanase},\ and\ \citenamefont {Ono}}]{Miyasaka2021}%
  \BibitemOpen
  \bibfield  {author} {\bibinfo {author} {\bibfnamefont {Y.}~\bibnamefont {Miyasaka}}, \bibinfo {author} {\bibfnamefont {R.}~\bibnamefont {Kawarazaki}}, \bibinfo {author} {\bibfnamefont {H.}~\bibnamefont {Narita}}, \bibinfo {author} {\bibfnamefont {F.}~\bibnamefont {Ando}}, \bibinfo {author} {\bibfnamefont {Y.}~\bibnamefont {Ikeda}}, \bibinfo {author} {\bibfnamefont {R.}~\bibnamefont {Hisatomi}}, \bibinfo {author} {\bibfnamefont {A.}~\bibnamefont {Daido}}, \bibinfo {author} {\bibfnamefont {Y.}~\bibnamefont {Shiota}}, \bibinfo {author} {\bibfnamefont {T.}~\bibnamefont {Moriyama}}, \bibinfo {author} {\bibfnamefont {Y.}~\bibnamefont {Yanase}},\ and\ \bibinfo {author} {\bibfnamefont {T.}~\bibnamefont {Ono}},\ }\bibfield  {title} {\bibinfo {title} {Observation of nonreciprocal superconducting critical field},\ }\href {https://doi.org/10.35848/1882-0786/ac03c0} {\bibfield  {journal} {\bibinfo  {journal} {Applied Physics Express}\ }\textbf {\bibinfo {volume} {14}},\ \bibinfo {pages} {073003} (\bibinfo {year}
  {2021})}\BibitemShut {NoStop}%
\bibitem [{\citenamefont {Karabassov}\ \emph {et~al.}(2022)\citenamefont {Karabassov}, \citenamefont {Bobkova}, \citenamefont {Golubov},\ and\ \citenamefont {Vasenko}}]{KarabassovBobkovaVasenko22}%
  \BibitemOpen
  \bibfield  {author} {\bibinfo {author} {\bibfnamefont {T.}~\bibnamefont {Karabassov}}, \bibinfo {author} {\bibfnamefont {I.~V.}\ \bibnamefont {Bobkova}}, \bibinfo {author} {\bibfnamefont {A.~A.}\ \bibnamefont {Golubov}},\ and\ \bibinfo {author} {\bibfnamefont {A.~S.}\ \bibnamefont {Vasenko}},\ }\bibfield  {title} {\bibinfo {title} {Hybrid helical state and superconducting diode effect in superconductor/ferromagnet/topological insulator heterostructures},\ }\href@noop {} {\bibfield  {journal} {\bibinfo  {journal} {Physical Review B}\ }\textbf {\bibinfo {volume} {106}},\ \bibinfo {pages} {224509} (\bibinfo {year} {2022})},\ \bibinfo {note} {publisher: American Physical Society}\BibitemShut {NoStop}%
\bibitem [{\citenamefont {Sundaresh}\ \emph {et~al.}(2023)\citenamefont {Sundaresh}, \citenamefont {V\"ayrynen}, \citenamefont {Lyanda-Geller},\ and\ \citenamefont {Rokhinson}}]{SundareshRokhinson23}%
  \BibitemOpen
  \bibfield  {author} {\bibinfo {author} {\bibfnamefont {A.}~\bibnamefont {Sundaresh}}, \bibinfo {author} {\bibfnamefont {J.~I.}\ \bibnamefont {V\"ayrynen}}, \bibinfo {author} {\bibfnamefont {Y.}~\bibnamefont {Lyanda-Geller}},\ and\ \bibinfo {author} {\bibfnamefont {L.~P.}\ \bibnamefont {Rokhinson}},\ }\bibfield  {title} {\bibinfo {title} {Diamagnetic mechanism of critical current non-reciprocity in multilayered superconductors},\ }\href@noop {} {\bibfield  {journal} {\bibinfo  {journal} {Nature Communications}\ }\textbf {\bibinfo {volume} {14}},\ \bibinfo {pages} {1628} (\bibinfo {year} {2023})},\ \bibinfo {note} {number: 1 Publisher: Nature Publishing Group}\BibitemShut {NoStop}%
\bibitem [{\citenamefont {Kealhofer}\ \emph {et~al.}(2023)\citenamefont {Kealhofer}, \citenamefont {Jeong}, \citenamefont {Rashidi}, \citenamefont {Balents},\ and\ \citenamefont {Stemmer}}]{KealhoferBalentsStemmer23}%
  \BibitemOpen
  \bibfield  {author} {\bibinfo {author} {\bibfnamefont {R.}~\bibnamefont {Kealhofer}}, \bibinfo {author} {\bibfnamefont {H.}~\bibnamefont {Jeong}}, \bibinfo {author} {\bibfnamefont {A.}~\bibnamefont {Rashidi}}, \bibinfo {author} {\bibfnamefont {L.}~\bibnamefont {Balents}},\ and\ \bibinfo {author} {\bibfnamefont {S.}~\bibnamefont {Stemmer}},\ }\bibfield  {title} {\bibinfo {title} {Anomalous superconducting diode effect in a polar superconductor},\ }\href@noop {} {\bibfield  {journal} {\bibinfo  {journal} {Physical Review B}\ }\textbf {\bibinfo {volume} {107}},\ \bibinfo {pages} {L100504} (\bibinfo {year} {2023})},\ \bibinfo {note} {publisher: American Physical Society}\BibitemShut {NoStop}%
\bibitem [{\citenamefont {Geng}\ \emph {et~al.}(2023)\citenamefont {Geng}, \citenamefont {Hijano}, \citenamefont {Ili\'c}, \citenamefont {Ilyn}, \citenamefont {Maasilta}, \citenamefont {Monfardini}, \citenamefont {Spies}, \citenamefont {Strambini}, \citenamefont {Virtanen}, \citenamefont {Calvo}, \citenamefont {Gonz\'alez-Orell\'ana}, \citenamefont {Helenius}, \citenamefont {Khorshidian}, \citenamefont {de~Araujo}, \citenamefont {Levy-Bertrand}, \citenamefont {Rogero}, \citenamefont {Giazotto}, \citenamefont {Bergeret},\ and\ \citenamefont {Heikkil\"a}}]{GengBergeretHeikkila23}%
  \BibitemOpen
  \bibfield  {author} {\bibinfo {author} {\bibfnamefont {Z.}~\bibnamefont {Geng}}, \bibinfo {author} {\bibfnamefont {A.}~\bibnamefont {Hijano}}, \bibinfo {author} {\bibfnamefont {S.}~\bibnamefont {Ili\'c}}, \bibinfo {author} {\bibfnamefont {M.}~\bibnamefont {Ilyn}}, \bibinfo {author} {\bibfnamefont {I.}~\bibnamefont {Maasilta}}, \bibinfo {author} {\bibfnamefont {A.}~\bibnamefont {Monfardini}}, \bibinfo {author} {\bibfnamefont {M.}~\bibnamefont {Spies}}, \bibinfo {author} {\bibfnamefont {E.}~\bibnamefont {Strambini}}, \bibinfo {author} {\bibfnamefont {P.}~\bibnamefont {Virtanen}}, \bibinfo {author} {\bibfnamefont {M.}~\bibnamefont {Calvo}}, \bibinfo {author} {\bibfnamefont {C.}~\bibnamefont {Gonz\'alez-Orell\'ana}}, \bibinfo {author} {\bibfnamefont {A.~P.}\ \bibnamefont {Helenius}}, \bibinfo {author} {\bibfnamefont {S.}~\bibnamefont {Khorshidian}}, \bibinfo {author} {\bibfnamefont {C.~I.~L.}\ \bibnamefont {de~Araujo}}, \bibinfo {author} {\bibfnamefont {F.}~\bibnamefont {Levy-Bertrand}}, \bibinfo {author}
  {\bibfnamefont {C.}~\bibnamefont {Rogero}}, \bibinfo {author} {\bibfnamefont {F.}~\bibnamefont {Giazotto}}, \bibinfo {author} {\bibfnamefont {F.~S.}\ \bibnamefont {Bergeret}},\ and\ \bibinfo {author} {\bibfnamefont {T.~T.}\ \bibnamefont {Heikkil\"a}},\ }\bibfield  {title} {\bibinfo {title} {Superconductor-ferromagnet hybrids for non-reciprocal electronics and detectors},\ }\href {https://doi.org/10.1088/1361-6668/ad01e9} {\bibfield  {journal} {\bibinfo  {journal} {Superconductor Science and Technology}\ }\textbf {\bibinfo {volume} {36}},\ \bibinfo {pages} {123001} (\bibinfo {year} {2023})}\BibitemShut {NoStop}%
\bibitem [{\citenamefont {Satchell}\ \emph {et~al.}(2023)\citenamefont {Satchell}, \citenamefont {Shepley}, \citenamefont {Rosamond},\ and\ \citenamefont {Burnell}}]{SatchellBurnell23}%
  \BibitemOpen
  \bibfield  {author} {\bibinfo {author} {\bibfnamefont {N.}~\bibnamefont {Satchell}}, \bibinfo {author} {\bibfnamefont {P.}~\bibnamefont {Shepley}}, \bibinfo {author} {\bibfnamefont {M.}~\bibnamefont {Rosamond}},\ and\ \bibinfo {author} {\bibfnamefont {G.}~\bibnamefont {Burnell}},\ }\bibfield  {title} {\bibinfo {title} {Supercurrent diode effect in thin film nb tracks},\ }\href {https://doi.org/10.1063/5.0141576} {\bibfield  {journal} {\bibinfo  {journal} {Journal of Applied Physics}\ }\textbf {\bibinfo {volume} {133}},\ \bibinfo {pages} {203901} (\bibinfo {year} {2023})},\ \Eprint {https://arxiv.org/abs/https://pubs.aip.org/aip/jap/article-pdf/doi/10.1063/5.0141576/18051614/203901\_1\_5.0141576.pdf} {https://pubs.aip.org/aip/jap/article-pdf/doi/10.1063/5.0141576/18051614/203901\_1\_5.0141576.pdf} \BibitemShut {NoStop}%
\bibitem [{\citenamefont {Putilov}\ \emph {et~al.}(2024)\citenamefont {Putilov}, \citenamefont {Mironov},\ and\ \citenamefont {Buzdin}}]{PutilovBuzdin24}%
  \BibitemOpen
  \bibfield  {author} {\bibinfo {author} {\bibfnamefont {A.~V.}\ \bibnamefont {Putilov}}, \bibinfo {author} {\bibfnamefont {S.~V.}\ \bibnamefont {Mironov}},\ and\ \bibinfo {author} {\bibfnamefont {A.~I.}\ \bibnamefont {Buzdin}},\ }\bibfield  {title} {\bibinfo {title} {Nonreciprocal electron transport in finite-size superconductor/ferromagnet bilayers with strong spin-orbit coupling},\ }\href@noop {} {\bibfield  {journal} {\bibinfo  {journal} {Physical Review B}\ }\textbf {\bibinfo {volume} {109}},\ \bibinfo {pages} {014510} (\bibinfo {year} {2024})},\ \bibinfo {note} {publisher: American Physical Society}\BibitemShut {NoStop}%
\bibitem [{\citenamefont {Vodolazov}\ and\ \citenamefont {Peeters}(2005)}]{VodolazovPeeters05}%
  \BibitemOpen
  \bibfield  {author} {\bibinfo {author} {\bibfnamefont {D.~Y.}\ \bibnamefont {Vodolazov}}\ and\ \bibinfo {author} {\bibfnamefont {F.~M.}\ \bibnamefont {Peeters}},\ }\bibfield  {title} {\bibinfo {title} {Superconducting rectifier based on the asymmetric surface barrier effect},\ }\href {https://doi.org/10.1103/PhysRevB.72.172508} {\bibfield  {journal} {\bibinfo  {journal} {Physical Review B}\ }\textbf {\bibinfo {volume} {72}},\ \bibinfo {pages} {172508} (\bibinfo {year} {2005})},\ \bibinfo {note} {publisher: American Physical Society}\BibitemShut {NoStop}%
\bibitem [{\citenamefont {Cerbu}\ \emph {et~al.}(2013)\citenamefont {Cerbu}, \citenamefont {Gladilin}, \citenamefont {Cuppens}, \citenamefont {Fritzsche}, \citenamefont {Tempere}, \citenamefont {Devreese}, \citenamefont {Moshchalkov}, \citenamefont {Silhanek},\ and\ \citenamefont {Vondel}}]{CerbuVondel13}%
  \BibitemOpen
  \bibfield  {author} {\bibinfo {author} {\bibfnamefont {D.}~\bibnamefont {Cerbu}}, \bibinfo {author} {\bibfnamefont {V.~N.}\ \bibnamefont {Gladilin}}, \bibinfo {author} {\bibfnamefont {J.}~\bibnamefont {Cuppens}}, \bibinfo {author} {\bibfnamefont {J.}~\bibnamefont {Fritzsche}}, \bibinfo {author} {\bibfnamefont {J.}~\bibnamefont {Tempere}}, \bibinfo {author} {\bibfnamefont {J.~T.}\ \bibnamefont {Devreese}}, \bibinfo {author} {\bibfnamefont {V.~V.}\ \bibnamefont {Moshchalkov}}, \bibinfo {author} {\bibfnamefont {A.~V.}\ \bibnamefont {Silhanek}},\ and\ \bibinfo {author} {\bibfnamefont {J.~V.~d.}\ \bibnamefont {Vondel}},\ }\bibfield  {title} {\bibinfo {title} {Vortex ratchet induced by controlled edge roughness},\ }\href {https://doi.org/10.1088/1367-2630/15/6/063022} {\bibfield  {journal} {\bibinfo  {journal} {New Journal of Physics}\ }\textbf {\bibinfo {volume} {15}},\ \bibinfo {pages} {063022} (\bibinfo {year} {2013})},\ \bibinfo {note} {publisher: IOP Publishing}\BibitemShut {NoStop}%
\bibitem [{\citenamefont {Gutfreund}\ \emph {et~al.}(2023)\citenamefont {Gutfreund}, \citenamefont {Matsuki}, \citenamefont {Plastovets}, \citenamefont {Noah}, \citenamefont {Gorzawski}, \citenamefont {Fridman}, \citenamefont {Yang}, \citenamefont {Buzdin}, \citenamefont {Millo}, \citenamefont {Robinson},\ and\ \citenamefont {Anahory}}]{GutfreundAnahory23}%
  \BibitemOpen
  \bibfield  {author} {\bibinfo {author} {\bibfnamefont {A.}~\bibnamefont {Gutfreund}}, \bibinfo {author} {\bibfnamefont {H.}~\bibnamefont {Matsuki}}, \bibinfo {author} {\bibfnamefont {V.}~\bibnamefont {Plastovets}}, \bibinfo {author} {\bibfnamefont {A.}~\bibnamefont {Noah}}, \bibinfo {author} {\bibfnamefont {L.}~\bibnamefont {Gorzawski}}, \bibinfo {author} {\bibfnamefont {N.}~\bibnamefont {Fridman}}, \bibinfo {author} {\bibfnamefont {G.}~\bibnamefont {Yang}}, \bibinfo {author} {\bibfnamefont {A.}~\bibnamefont {Buzdin}}, \bibinfo {author} {\bibfnamefont {O.}~\bibnamefont {Millo}}, \bibinfo {author} {\bibfnamefont {J.~W.~A.}\ \bibnamefont {Robinson}},\ and\ \bibinfo {author} {\bibfnamefont {Y.}~\bibnamefont {Anahory}},\ }\bibfield  {title} {\bibinfo {title} {Direct observation of a superconducting vortex diode},\ }\href@noop {} {\bibfield  {journal} {\bibinfo  {journal} {Nature Communications}\ }\textbf {\bibinfo {volume} {14}},\ \bibinfo {pages} {1630} (\bibinfo {year} {2023})},\ \bibinfo {note} {number: 1
  Publisher: Nature Publishing Group}\BibitemShut {NoStop}%
\bibitem [{\citenamefont {Hou}\ \emph {et~al.}(2023)\citenamefont {Hou}, \citenamefont {Nichele}, \citenamefont {Chi}, \citenamefont {Lodesani}, \citenamefont {Wu}, \citenamefont {Ritter}, \citenamefont {Haxell}, \citenamefont {Davydova}, \citenamefont {Ili\ifmmode~\acute{c}\else \'{c}\fi{}}, \citenamefont {Glezakou-Elbert}, \citenamefont {Varambally}, \citenamefont {Bergeret}, \citenamefont {Kamra}, \citenamefont {Fu}, \citenamefont {Lee},\ and\ \citenamefont {Moodera}}]{HouMoodera23}%
  \BibitemOpen
  \bibfield  {author} {\bibinfo {author} {\bibfnamefont {Y.}~\bibnamefont {Hou}}, \bibinfo {author} {\bibfnamefont {F.}~\bibnamefont {Nichele}}, \bibinfo {author} {\bibfnamefont {H.}~\bibnamefont {Chi}}, \bibinfo {author} {\bibfnamefont {A.}~\bibnamefont {Lodesani}}, \bibinfo {author} {\bibfnamefont {Y.}~\bibnamefont {Wu}}, \bibinfo {author} {\bibfnamefont {M.~F.}\ \bibnamefont {Ritter}}, \bibinfo {author} {\bibfnamefont {D.~Z.}\ \bibnamefont {Haxell}}, \bibinfo {author} {\bibfnamefont {M.}~\bibnamefont {Davydova}}, \bibinfo {author} {\bibfnamefont {S.}~\bibnamefont {Ili\ifmmode~\acute{c}\else \'{c}\fi{}}}, \bibinfo {author} {\bibfnamefont {O.}~\bibnamefont {Glezakou-Elbert}}, \bibinfo {author} {\bibfnamefont {A.}~\bibnamefont {Varambally}}, \bibinfo {author} {\bibfnamefont {F.~S.}\ \bibnamefont {Bergeret}}, \bibinfo {author} {\bibfnamefont {A.}~\bibnamefont {Kamra}}, \bibinfo {author} {\bibfnamefont {L.}~\bibnamefont {Fu}}, \bibinfo {author} {\bibfnamefont {P.~A.}\ \bibnamefont {Lee}},\ and\ \bibinfo
  {author} {\bibfnamefont {J.~S.}\ \bibnamefont {Moodera}},\ }\href@noop {} {\bibinfo {title} {Ubiquitous superconducting diode effect in superconductor thin films}} (\bibinfo {year} {2023})\BibitemShut {NoStop}%
\bibitem [{\citenamefont {Lyu}\ \emph {et~al.}(2021)\citenamefont {Lyu}, \citenamefont {Jiang}, \citenamefont {Wang}, \citenamefont {Xiao}, \citenamefont {Dong}, \citenamefont {Chen}, \citenamefont {Milosevic}, \citenamefont {Wang}, \citenamefont {Divan}, \citenamefont {Pearson}, \citenamefont {Wu}, \citenamefont {Peeters},\ and\ \citenamefont {Kwok}}]{LyuKwok21}%
  \BibitemOpen
  \bibfield  {author} {\bibinfo {author} {\bibfnamefont {Y.-Y.}\ \bibnamefont {Lyu}}, \bibinfo {author} {\bibfnamefont {J.}~\bibnamefont {Jiang}}, \bibinfo {author} {\bibfnamefont {Y.-L.}\ \bibnamefont {Wang}}, \bibinfo {author} {\bibfnamefont {Z.-L.}\ \bibnamefont {Xiao}}, \bibinfo {author} {\bibfnamefont {S.}~\bibnamefont {Dong}}, \bibinfo {author} {\bibfnamefont {Q.-H.}\ \bibnamefont {Chen}}, \bibinfo {author} {\bibfnamefont {M.~V.}\ \bibnamefont {Milosevic}}, \bibinfo {author} {\bibfnamefont {H.}~\bibnamefont {Wang}}, \bibinfo {author} {\bibfnamefont {R.}~\bibnamefont {Divan}}, \bibinfo {author} {\bibfnamefont {J.~E.}\ \bibnamefont {Pearson}}, \bibinfo {author} {\bibfnamefont {P.}~\bibnamefont {Wu}}, \bibinfo {author} {\bibfnamefont {F.~M.}\ \bibnamefont {Peeters}},\ and\ \bibinfo {author} {\bibfnamefont {W.-K.}\ \bibnamefont {Kwok}},\ }\bibfield  {title} {\bibinfo {title} {Superconducting diode effect via conformal-mapped nanoholes},\ }\href@noop {} {\bibfield  {journal} {\bibinfo  {journal} {Nature
  Communications}\ }\textbf {\bibinfo {volume} {12}},\ \bibinfo {pages} {2703} (\bibinfo {year} {2021})},\ \bibinfo {note} {number: 1 Publisher: Nature Publishing Group}\BibitemShut {NoStop}%
\bibitem [{\citenamefont {Wakatsuki}\ and\ \citenamefont {Nagaosa}(2018)}]{WakatsukiNagaosa18}%
  \BibitemOpen
  \bibfield  {author} {\bibinfo {author} {\bibfnamefont {R.}~\bibnamefont {Wakatsuki}}\ and\ \bibinfo {author} {\bibfnamefont {N.}~\bibnamefont {Nagaosa}},\ }\bibfield  {title} {\bibinfo {title} {Nonreciprocal {Current} in {Noncentrosymmetric} {Rashba} {Superconductors}},\ }\href@noop {} {\bibfield  {journal} {\bibinfo  {journal} {Physical Review Letters}\ }\textbf {\bibinfo {volume} {121}},\ \bibinfo {pages} {026601} (\bibinfo {year} {2018})},\ \bibinfo {note} {publisher: American Physical Society}\BibitemShut {NoStop}%
\bibitem [{\citenamefont {Hoshino}\ \emph {et~al.}(2018)\citenamefont {Hoshino}, \citenamefont {Wakatsuki}, \citenamefont {Hamamoto},\ and\ \citenamefont {Nagaosa}}]{HoshinoNagaosa18}%
  \BibitemOpen
  \bibfield  {author} {\bibinfo {author} {\bibfnamefont {S.}~\bibnamefont {Hoshino}}, \bibinfo {author} {\bibfnamefont {R.}~\bibnamefont {Wakatsuki}}, \bibinfo {author} {\bibfnamefont {K.}~\bibnamefont {Hamamoto}},\ and\ \bibinfo {author} {\bibfnamefont {N.}~\bibnamefont {Nagaosa}},\ }\bibfield  {title} {\bibinfo {title} {Nonreciprocal charge transport in two-dimensional noncentrosymmetric superconductors},\ }\href@noop {} {\bibfield  {journal} {\bibinfo  {journal} {Physical Review B}\ }\textbf {\bibinfo {volume} {98}},\ \bibinfo {pages} {054510} (\bibinfo {year} {2018})},\ \bibinfo {note} {publisher: American Physical Society}\BibitemShut {NoStop}%
\bibitem [{\citenamefont {Daido}\ \emph {et~al.}(2022)\citenamefont {Daido}, \citenamefont {Ikeda},\ and\ \citenamefont {Yanase}}]{DaidoYanase22}%
  \BibitemOpen
  \bibfield  {author} {\bibinfo {author} {\bibfnamefont {A.}~\bibnamefont {Daido}}, \bibinfo {author} {\bibfnamefont {Y.}~\bibnamefont {Ikeda}},\ and\ \bibinfo {author} {\bibfnamefont {Y.}~\bibnamefont {Yanase}},\ }\bibfield  {title} {\bibinfo {title} {Intrinsic {Superconducting} {Diode} {Effect}},\ }\href@noop {} {\bibfield  {journal} {\bibinfo  {journal} {Physical Review Letters}\ }\textbf {\bibinfo {volume} {128}},\ \bibinfo {pages} {037001} (\bibinfo {year} {2022})},\ \bibinfo {note} {publisher: American Physical Society}\BibitemShut {NoStop}%
\bibitem [{\citenamefont {Daido}\ and\ \citenamefont {Yanase}(2022)}]{DaidoYanase22_2}%
  \BibitemOpen
  \bibfield  {author} {\bibinfo {author} {\bibfnamefont {A.}~\bibnamefont {Daido}}\ and\ \bibinfo {author} {\bibfnamefont {Y.}~\bibnamefont {Yanase}},\ }\bibfield  {title} {\bibinfo {title} {Superconducting diode effect and nonreciprocal transition lines},\ }\href {https://doi.org/10.1103/PhysRevB.106.205206} {\bibfield  {journal} {\bibinfo  {journal} {Phys. Rev. B}\ }\textbf {\bibinfo {volume} {106}},\ \bibinfo {pages} {205206} (\bibinfo {year} {2022})}\BibitemShut {NoStop}%
\bibitem [{\citenamefont {Yuan}\ and\ \citenamefont {Fu}(2022)}]{YuanFu22}%
  \BibitemOpen
  \bibfield  {author} {\bibinfo {author} {\bibfnamefont {N.~F.~Q.}\ \bibnamefont {Yuan}}\ and\ \bibinfo {author} {\bibfnamefont {L.}~\bibnamefont {Fu}},\ }\bibfield  {title} {\bibinfo {title} {Supercurrent diode effect and finite-momentum superconductors},\ }\href@noop {} {\bibfield  {journal} {\bibinfo  {journal} {Proceedings of the National Academy of Sciences}\ }\textbf {\bibinfo {volume} {119}},\ \bibinfo {pages} {e2119548119} (\bibinfo {year} {2022})},\ \bibinfo {note} {publisher: Proceedings of the National Academy of Sciences}\BibitemShut {NoStop}%
\bibitem [{\citenamefont {He}\ \emph {et~al.}(2022)\citenamefont {He}, \citenamefont {Tanaka},\ and\ \citenamefont {Nagaosa}}]{HeNagaosa22}%
  \BibitemOpen
  \bibfield  {author} {\bibinfo {author} {\bibfnamefont {J.~J.}\ \bibnamefont {He}}, \bibinfo {author} {\bibfnamefont {Y.}~\bibnamefont {Tanaka}},\ and\ \bibinfo {author} {\bibfnamefont {N.}~\bibnamefont {Nagaosa}},\ }\bibfield  {title} {\bibinfo {title} {A phenomenological theory of superconductor diodes},\ }\href@noop {} {\bibfield  {journal} {\bibinfo  {journal} {New Journal of Physics}\ }\textbf {\bibinfo {volume} {24}},\ \bibinfo {pages} {053014} (\bibinfo {year} {2022})},\ \bibinfo {note} {publisher: IOP Publishing}\BibitemShut {NoStop}%
\bibitem [{\citenamefont {Ilic}\ and\ \citenamefont {Bergeret}(2022)}]{IlicBergeret22}%
  \BibitemOpen
  \bibfield  {author} {\bibinfo {author} {\bibfnamefont {S.}~\bibnamefont {Ilic}}\ and\ \bibinfo {author} {\bibfnamefont {F.~S.}\ \bibnamefont {Bergeret}},\ }\bibfield  {title} {\bibinfo {title} {{Theory of the Supercurrent Diode Effect in {R}ashba Superconductors with Arbitrary Disorder}},\ }\href {https://doi.org/10.1103/PhysRevLett.128.177001} {\bibfield  {journal} {\bibinfo  {journal} {Phys. Rev. Lett.}\ }\textbf {\bibinfo {volume} {128}},\ \bibinfo {pages} {177001} (\bibinfo {year} {2022})}\BibitemShut {NoStop}%
\bibitem [{\citenamefont {Bychkov}\ and\ \citenamefont {Rashba}(1984)}]{Rashba1984}%
  \BibitemOpen
  \bibfield  {author} {\bibinfo {author} {\bibfnamefont {Y.~A.}\ \bibnamefont {Bychkov}}\ and\ \bibinfo {author} {\bibfnamefont {E.~I.}\ \bibnamefont {Rashba}},\ }\bibfield  {title} {\bibinfo {title} {Properties of a 2{D} electron gas with lifted spectral degeneracy},\ }\href@noop {} {\bibfield  {journal} {\bibinfo  {journal} {Sov. Phys. JETP Lett.}\ }\textbf {\bibinfo {volume} {39}},\ \bibinfo {pages} {78} (\bibinfo {year} {1984})}\BibitemShut {NoStop}%
\bibitem [{\citenamefont {Lu}\ \emph {et~al.}(2015)\citenamefont {Lu}, \citenamefont {Zheliuk}, \citenamefont {Leermakers}, \citenamefont {Yuan}, \citenamefont {Zeitler}, \citenamefont {Law},\ and\ \citenamefont {Ye}}]{lu2015evidence}%
  \BibitemOpen
  \bibfield  {author} {\bibinfo {author} {\bibfnamefont {J.~M.}\ \bibnamefont {Lu}}, \bibinfo {author} {\bibfnamefont {O.}~\bibnamefont {Zheliuk}}, \bibinfo {author} {\bibfnamefont {I.}~\bibnamefont {Leermakers}}, \bibinfo {author} {\bibfnamefont {N.~F.~Q.}\ \bibnamefont {Yuan}}, \bibinfo {author} {\bibfnamefont {U.}~\bibnamefont {Zeitler}}, \bibinfo {author} {\bibfnamefont {K.~T.}\ \bibnamefont {Law}},\ and\ \bibinfo {author} {\bibfnamefont {J.~T.}\ \bibnamefont {Ye}},\ }\bibfield  {title} {\bibinfo {title} {{Evidence for two-dimensional Ising superconductivity in gated MoS$_2$}},\ }\href {https://doi.org/10.1126/science.aab2277} {\bibfield  {journal} {\bibinfo  {journal} {Science}\ }\textbf {\bibinfo {volume} {350}},\ \bibinfo {pages} {1353} (\bibinfo {year} {2015})}\BibitemShut {NoStop}%
\bibitem [{\citenamefont {Saito}\ \emph {et~al.}(2016)\citenamefont {Saito}, \citenamefont {Nakamura}, \citenamefont {Bahramy}, \citenamefont {Kohama}, \citenamefont {Ye}, \citenamefont {Kasahara}, \citenamefont {Nakagawa}, \citenamefont {Onga}, \citenamefont {Tokunaga}, \citenamefont {Nojima}, \citenamefont {Yanase},\ and\ \citenamefont {Iwasa}}]{saito2016superconductivity}%
  \BibitemOpen
  \bibfield  {author} {\bibinfo {author} {\bibfnamefont {Y.}~\bibnamefont {Saito}}, \bibinfo {author} {\bibfnamefont {Y.}~\bibnamefont {Nakamura}}, \bibinfo {author} {\bibfnamefont {M.~S.}\ \bibnamefont {Bahramy}}, \bibinfo {author} {\bibfnamefont {Y.}~\bibnamefont {Kohama}}, \bibinfo {author} {\bibfnamefont {J.}~\bibnamefont {Ye}}, \bibinfo {author} {\bibfnamefont {Y.}~\bibnamefont {Kasahara}}, \bibinfo {author} {\bibfnamefont {Y.}~\bibnamefont {Nakagawa}}, \bibinfo {author} {\bibfnamefont {M.}~\bibnamefont {Onga}}, \bibinfo {author} {\bibfnamefont {M.}~\bibnamefont {Tokunaga}}, \bibinfo {author} {\bibfnamefont {T.}~\bibnamefont {Nojima}}, \bibinfo {author} {\bibfnamefont {Y.}~\bibnamefont {Yanase}},\ and\ \bibinfo {author} {\bibfnamefont {Y.}~\bibnamefont {Iwasa}},\ }\bibfield  {title} {\bibinfo {title} {{Superconductivity protected by spin-valley locking in ion-gated MoS$_2$}},\ }\href {https://doi.org/10.1038/nphys3580} {\bibfield  {journal} {\bibinfo  {journal} {Nat. Phys.}\ }\textbf {\bibinfo {volume}
  {12}},\ \bibinfo {pages} {144} (\bibinfo {year} {2016})}\BibitemShut {NoStop}%
\bibitem [{\citenamefont {Xi}\ \emph {et~al.}(2016)\citenamefont {Xi}, \citenamefont {Wang}, \citenamefont {Zhao}, \citenamefont {Park}, \citenamefont {Law}, \citenamefont {Berger}, \citenamefont {Forr{\'{o}}}, \citenamefont {Shan},\ and\ \citenamefont {Mak}}]{xi2016Ising}%
  \BibitemOpen
  \bibfield  {author} {\bibinfo {author} {\bibfnamefont {X.}~\bibnamefont {Xi}}, \bibinfo {author} {\bibfnamefont {Z.}~\bibnamefont {Wang}}, \bibinfo {author} {\bibfnamefont {W.}~\bibnamefont {Zhao}}, \bibinfo {author} {\bibfnamefont {J.-H.}\ \bibnamefont {Park}}, \bibinfo {author} {\bibfnamefont {K.~T.}\ \bibnamefont {Law}}, \bibinfo {author} {\bibfnamefont {H.}~\bibnamefont {Berger}}, \bibinfo {author} {\bibfnamefont {L.}~\bibnamefont {Forr{\'{o}}}}, \bibinfo {author} {\bibfnamefont {J.}~\bibnamefont {Shan}},\ and\ \bibinfo {author} {\bibfnamefont {K.~F.}\ \bibnamefont {Mak}},\ }\bibfield  {title} {\bibinfo {title} {{Ising pairing in superconducting NbSe$_2$ atomic layers}},\ }\href {https://doi.org/10.1038/nphys3538} {\bibfield  {journal} {\bibinfo  {journal} {Nat. Phys.}\ }\textbf {\bibinfo {volume} {12}},\ \bibinfo {pages} {139} (\bibinfo {year} {2016})}\BibitemShut {NoStop}%
\bibitem [{\citenamefont {Costanzo}\ \emph {et~al.}(2018)\citenamefont {Costanzo}, \citenamefont {Zhang}, \citenamefont {Reddy}, \citenamefont {Berger},\ and\ \citenamefont {Morpurgo}}]{costanzo2018tunnelling}%
  \BibitemOpen
  \bibfield  {author} {\bibinfo {author} {\bibfnamefont {D.}~\bibnamefont {Costanzo}}, \bibinfo {author} {\bibfnamefont {H.}~\bibnamefont {Zhang}}, \bibinfo {author} {\bibfnamefont {B.~A.}\ \bibnamefont {Reddy}}, \bibinfo {author} {\bibfnamefont {H.}~\bibnamefont {Berger}},\ and\ \bibinfo {author} {\bibfnamefont {A.~F.}\ \bibnamefont {Morpurgo}},\ }\bibfield  {title} {\bibinfo {title} {{Tunnelling spectroscopy of gate-induced superconductivity in MoS$_2$}},\ }\href {https://doi.org/10.1038/s41565-018-0122-2} {\bibfield  {journal} {\bibinfo  {journal} {Nat. Nanotechnol.}\ }\textbf {\bibinfo {volume} {13}},\ \bibinfo {pages} {483} (\bibinfo {year} {2018})}\BibitemShut {NoStop}%
\bibitem [{\citenamefont {de~la Barrera}\ \emph {et~al.}(2018)\citenamefont {de~la Barrera}, \citenamefont {Sinko}, \citenamefont {Gopalan}, \citenamefont {Sivadas}, \citenamefont {Seyler}, \citenamefont {Watanabe}, \citenamefont {Taniguchi}, \citenamefont {Tsen}, \citenamefont {Xu}, \citenamefont {Xiao},\ and\ \citenamefont {Hunt}}]{delaBarrera2018}%
  \BibitemOpen
  \bibfield  {author} {\bibinfo {author} {\bibfnamefont {S.~C.}\ \bibnamefont {de~la Barrera}}, \bibinfo {author} {\bibfnamefont {M.~R.}\ \bibnamefont {Sinko}}, \bibinfo {author} {\bibfnamefont {D.~P.}\ \bibnamefont {Gopalan}}, \bibinfo {author} {\bibfnamefont {N.}~\bibnamefont {Sivadas}}, \bibinfo {author} {\bibfnamefont {K.~L.}\ \bibnamefont {Seyler}}, \bibinfo {author} {\bibfnamefont {K.}~\bibnamefont {Watanabe}}, \bibinfo {author} {\bibfnamefont {T.}~\bibnamefont {Taniguchi}}, \bibinfo {author} {\bibfnamefont {A.~W.}\ \bibnamefont {Tsen}}, \bibinfo {author} {\bibfnamefont {X.}~\bibnamefont {Xu}}, \bibinfo {author} {\bibfnamefont {D.}~\bibnamefont {Xiao}},\ and\ \bibinfo {author} {\bibfnamefont {B.~M.}\ \bibnamefont {Hunt}},\ }\bibfield  {title} {\bibinfo {title} {{Tuning Ising superconductivity with layer and spin-orbit coupling in two-dimensional transition-metal dichalcogenides}},\ }\href {https://doi.org/10.1038/s41467-018-03888-4} {\bibfield  {journal} {\bibinfo  {journal} {Nat. Commun.}\ }\textbf
  {\bibinfo {volume} {9}},\ \bibinfo {pages} {1427} (\bibinfo {year} {2018})}\BibitemShut {NoStop}%
\bibitem [{\citenamefont {Houzet}\ and\ \citenamefont {Meyer}(2015)}]{Houzet2015}%
  \BibitemOpen
  \bibfield  {author} {\bibinfo {author} {\bibfnamefont {M.}~\bibnamefont {Houzet}}\ and\ \bibinfo {author} {\bibfnamefont {J.~S.}\ \bibnamefont {Meyer}},\ }\bibfield  {title} {\bibinfo {title} {Quasiclassical theory of disordered rashba superconductors},\ }\href {https://doi.org/10.1103/PhysRevB.92.014509} {\bibfield  {journal} {\bibinfo  {journal} {Phys. Rev. B}\ }\textbf {\bibinfo {volume} {92}},\ \bibinfo {pages} {014509} (\bibinfo {year} {2015})}\BibitemShut {NoStop}%
\bibitem [{\citenamefont {Haim}\ \emph {et~al.}(2022)\citenamefont {Haim}, \citenamefont {Levchenko},\ and\ \citenamefont {Khodas}}]{Haim2022}%
  \BibitemOpen
  \bibfield  {author} {\bibinfo {author} {\bibfnamefont {M.}~\bibnamefont {Haim}}, \bibinfo {author} {\bibfnamefont {A.}~\bibnamefont {Levchenko}},\ and\ \bibinfo {author} {\bibfnamefont {M.}~\bibnamefont {Khodas}},\ }\bibfield  {title} {\bibinfo {title} {Mechanisms of in-plane magnetic anisotropy in superconducting ${\mathrm{nbse}}_{2}$},\ }\href {https://doi.org/10.1103/PhysRevB.105.024515} {\bibfield  {journal} {\bibinfo  {journal} {Phys. Rev. B}\ }\textbf {\bibinfo {volume} {105}},\ \bibinfo {pages} {024515} (\bibinfo {year} {2022})}\BibitemShut {NoStop}%
\bibitem [{\citenamefont {Hasan}\ \emph {et~al.}(2024)\citenamefont {Hasan}, \citenamefont {Shaffer}, \citenamefont {Khodas},\ and\ \citenamefont {Levchenko}}]{Hasan2024}%
  \BibitemOpen
  \bibfield  {author} {\bibinfo {author} {\bibfnamefont {J.}~\bibnamefont {Hasan}}, \bibinfo {author} {\bibfnamefont {D.}~\bibnamefont {Shaffer}}, \bibinfo {author} {\bibfnamefont {M.}~\bibnamefont {Khodas}},\ and\ \bibinfo {author} {\bibfnamefont {A.}~\bibnamefont {Levchenko}},\ }\bibfield  {title} {\bibinfo {title} {Supercurrent diode effect in helical superconductors},\ }\href {https://doi.org/10.1103/PhysRevB.110.024508} {\bibfield  {journal} {\bibinfo  {journal} {Phys. Rev. B}\ }\textbf {\bibinfo {volume} {110}},\ \bibinfo {pages} {024508} (\bibinfo {year} {2024})}\BibitemShut {NoStop}%
\bibitem [{\citenamefont {Hatano}\ \emph {et~al.}(2007)\citenamefont {Hatano}, \citenamefont {Shirasaki},\ and\ \citenamefont {Nakamura}}]{Hatano2007}%
  \BibitemOpen
  \bibfield  {author} {\bibinfo {author} {\bibfnamefont {N.}~\bibnamefont {Hatano}}, \bibinfo {author} {\bibfnamefont {R.~b.~o.}\ \bibnamefont {Shirasaki}},\ and\ \bibinfo {author} {\bibfnamefont {H.}~\bibnamefont {Nakamura}},\ }\bibfield  {title} {\bibinfo {title} {Non-{A}belian gauge field theory of the spin-orbit interaction and a perfect spin filter},\ }\href {https://doi.org/10.1103/PhysRevA.75.032107} {\bibfield  {journal} {\bibinfo  {journal} {Phys. Rev. A}\ }\textbf {\bibinfo {volume} {75}},\ \bibinfo {pages} {032107} (\bibinfo {year} {2007})}\BibitemShut {NoStop}%
\bibitem [{\citenamefont {Tokatly}(2008)}]{Tokatly2008}%
  \BibitemOpen
  \bibfield  {author} {\bibinfo {author} {\bibfnamefont {I.~V.}\ \bibnamefont {Tokatly}},\ }\bibfield  {title} {\bibinfo {title} {Equilibrium spin currents: Non-{A}belian gauge invariance and color diamagnetism in condensed matter},\ }\href {https://doi.org/10.1103/PhysRevLett.101.106601} {\bibfield  {journal} {\bibinfo  {journal} {Phys. Rev. Lett.}\ }\textbf {\bibinfo {volume} {101}},\ \bibinfo {pages} {106601} (\bibinfo {year} {2008})}\BibitemShut {NoStop}%
\bibitem [{\citenamefont {Kaplan}(1983)}]{Kaplan1983}%
  \BibitemOpen
  \bibfield  {author} {\bibinfo {author} {\bibfnamefont {T.~A.}\ \bibnamefont {Kaplan}},\ }\bibfield  {title} {\bibinfo {title} {Single-band {H}ubbard model with spin-orbit coupling},\ }\href {https://doi.org/10.1007/BF01301591} {\bibfield  {journal} {\bibinfo  {journal} {Zeitschrift f{\"u}r Physik B Condensed Matter}\ }\textbf {\bibinfo {volume} {49}},\ \bibinfo {pages} {313} (\bibinfo {year} {1983})}\BibitemShut {NoStop}%
\bibitem [{\citenamefont {Avishai}\ \emph {et~al.}(2019)\citenamefont {Avishai}, \citenamefont {Totsuka},\ and\ \citenamefont {Nagaosa}}]{Avishai2019}%
  \BibitemOpen
  \bibfield  {author} {\bibinfo {author} {\bibfnamefont {Y.}~\bibnamefont {Avishai}}, \bibinfo {author} {\bibfnamefont {K.}~\bibnamefont {Totsuka}},\ and\ \bibinfo {author} {\bibfnamefont {N.}~\bibnamefont {Nagaosa}},\ }\bibfield  {title} {\bibinfo {title} {Non-{A}belian {A}haronov-{C}asher phase factor in mesoscopic systems},\ }\href {https://doi.org/10.7566/JPSJ.88.084705} {\bibfield  {journal} {\bibinfo  {journal} {Journal of the Physical Society of Japan}\ }\textbf {\bibinfo {volume} {88}},\ \bibinfo {pages} {084705} (\bibinfo {year} {2019})},\ \Eprint {https://arxiv.org/abs/https://doi.org/10.7566/JPSJ.88.084705} {https://doi.org/10.7566/JPSJ.88.084705} \BibitemShut {NoStop}%
\bibitem [{\citenamefont {Chen}\ and\ \citenamefont {Chang}(2008)}]{Chen2008}%
  \BibitemOpen
  \bibfield  {author} {\bibinfo {author} {\bibfnamefont {S.-H.}\ \bibnamefont {Chen}}\ and\ \bibinfo {author} {\bibfnamefont {C.-R.}\ \bibnamefont {Chang}},\ }\bibfield  {title} {\bibinfo {title} {Non-{A}belian spin-orbit gauge: Persistent spin helix and quantum square ring},\ }\href {https://doi.org/10.1103/PhysRevB.77.045324} {\bibfield  {journal} {\bibinfo  {journal} {Phys. Rev. B}\ }\textbf {\bibinfo {volume} {77}},\ \bibinfo {pages} {045324} (\bibinfo {year} {2008})}\BibitemShut {NoStop}%
\bibitem [{\citenamefont {Tokatly}\ and\ \citenamefont {Sherman}(2010)}]{Tokatly2010}%
  \BibitemOpen
  \bibfield  {author} {\bibinfo {author} {\bibfnamefont {I.}~\bibnamefont {Tokatly}}\ and\ \bibinfo {author} {\bibfnamefont {E.}~\bibnamefont {Sherman}},\ }\bibfield  {title} {\bibinfo {title} {Gauge theory approach for diffusive and precessional spin dynamics in a two-dimensional electron gas},\ }\href {https://doi.org/https://doi.org/10.1016/j.aop.2010.01.007} {\bibfield  {journal} {\bibinfo  {journal} {Annals of Physics}\ }\textbf {\bibinfo {volume} {325}},\ \bibinfo {pages} {1104} (\bibinfo {year} {2010})}\BibitemShut {NoStop}%
\bibitem [{\citenamefont {Kotetes}\ \emph {et~al.}(2023)\citenamefont {Kotetes}, \citenamefont {Sura},\ and\ \citenamefont {Andersen}}]{KotetesAndersen23}%
  \BibitemOpen
  \bibfield  {author} {\bibinfo {author} {\bibfnamefont {P.}~\bibnamefont {Kotetes}}, \bibinfo {author} {\bibfnamefont {H.~O.~M.}\ \bibnamefont {Sura}},\ and\ \bibinfo {author} {\bibfnamefont {B.~M.}\ \bibnamefont {Andersen}},\ }\bibfield  {title} {\bibinfo {title} {{Anatomy of spin and current generation from magnetization gradients in topological insulators and Rashba metals}},\ }\href@noop {} {\bibfield  {journal} {\bibinfo  {journal} {Physical Review B}\ }\textbf {\bibinfo {volume} {108}},\ \bibinfo {pages} {155310} (\bibinfo {year} {2023})},\ \bibinfo {note} {publisher: American Physical Society}\BibitemShut {NoStop}%
\bibitem [{\citenamefont {Roig}\ \emph {et~al.}(2024)\citenamefont {Roig}, \citenamefont {Kotetes},\ and\ \citenamefont {Andersen}}]{RoigAndersen24}%
  \BibitemOpen
  \bibfield  {author} {\bibinfo {author} {\bibfnamefont {M.}~\bibnamefont {Roig}}, \bibinfo {author} {\bibfnamefont {P.}~\bibnamefont {Kotetes}},\ and\ \bibinfo {author} {\bibfnamefont {B.~M.}\ \bibnamefont {Andersen}},\ }\bibfield  {title} {\bibinfo {title} {Superconducting diodes from magnetization gradients},\ }\href {https://doi.org/10.1103/PhysRevB.109.144503} {\bibfield  {journal} {\bibinfo  {journal} {Physical Review B}\ }\textbf {\bibinfo {volume} {109}},\ \bibinfo {pages} {144503} (\bibinfo {year} {2024})},\ \bibinfo {note} {publisher: American Physical Society}\BibitemShut {NoStop}%
\bibitem [{\citenamefont {Tinkham}(2004)}]{Tinkham2004}%
  \BibitemOpen
  \bibfield  {author} {\bibinfo {author} {\bibfnamefont {M.}~\bibnamefont {Tinkham}},\ }\href {http://www.worldcat.org/isbn/0486435032} {\emph {\bibinfo {title} {Introduction to Superconductivity}}},\ \bibinfo {edition} {2nd}\ ed.\ (\bibinfo  {publisher} {Dover Publications},\ \bibinfo {year} {2004})\BibitemShut {NoStop}%
\bibitem [{\citenamefont {Edelstein}(2021)}]{edelstein_ginzburg-landau_2021}%
  \BibitemOpen
  \bibfield  {author} {\bibinfo {author} {\bibfnamefont {V.~M.}\ \bibnamefont {Edelstein}},\ }\bibfield  {title} {\bibinfo {title} {Ginzburg-{Landau} theory for impure superconductors of polar symmetry},\ }\href@noop {} {\bibfield  {journal} {\bibinfo  {journal} {Physical Review B}\ }\textbf {\bibinfo {volume} {103}},\ \bibinfo {pages} {094507} (\bibinfo {year} {2021})},\ \bibinfo {note} {publisher: American Physical Society}\BibitemShut {NoStop}%
\bibitem [{\citenamefont {Agterberg}(2012)}]{Agterberg2012}%
  \BibitemOpen
  \bibfield  {author} {\bibinfo {author} {\bibfnamefont {D.~F.}\ \bibnamefont {Agterberg}},\ }\bibinfo {title} {Magnetoelectric effects, helical phases, and fflo phases},\ in\ \href {https://doi.org/10.1007/978-3-642-24624-1_5} {\emph {\bibinfo {booktitle} {Non-Centrosymmetric Superconductors: Introduction and Overview}}},\ \bibinfo {editor} {edited by\ \bibinfo {editor} {\bibfnamefont {E.}~\bibnamefont {Bauer}}\ and\ \bibinfo {editor} {\bibfnamefont {M.}~\bibnamefont {Sigrist}}}\ (\bibinfo  {publisher} {Springer Berlin Heidelberg},\ \bibinfo {address} {Berlin, Heidelberg},\ \bibinfo {year} {2012})\ pp.\ \bibinfo {pages} {155--170}\BibitemShut {NoStop}%
\bibitem [{\citenamefont {Smidman}\ \emph {et~al.}(2017)\citenamefont {Smidman}, \citenamefont {Salamon}, \citenamefont {Yuan},\ and\ \citenamefont {Agterberg}}]{SmidmanAgterberg17}%
  \BibitemOpen
  \bibfield  {author} {\bibinfo {author} {\bibfnamefont {M.}~\bibnamefont {Smidman}}, \bibinfo {author} {\bibfnamefont {M.~B.}\ \bibnamefont {Salamon}}, \bibinfo {author} {\bibfnamefont {H.~Q.}\ \bibnamefont {Yuan}},\ and\ \bibinfo {author} {\bibfnamefont {D.~F.}\ \bibnamefont {Agterberg}},\ }\bibfield  {title} {\bibinfo {title} {Superconductivity and spin-orbit coupling in non-centrosymmetric materials: a review},\ }\href@noop {} {\bibfield  {journal} {\bibinfo  {journal} {Reports on Progress in Physics}\ }\textbf {\bibinfo {volume} {80}},\ \bibinfo {pages} {036501} (\bibinfo {year} {2017})},\ \bibinfo {note} {publisher: IOP Publishing}\BibitemShut {NoStop}%
\bibitem [{\citenamefont {Mineev}\ and\ \citenamefont {Samokhin}(1994)}]{MineevSamokhin94}%
  \BibitemOpen
  \bibfield  {author} {\bibinfo {author} {\bibfnamefont {V.~P.}\ \bibnamefont {Mineev}}\ and\ \bibinfo {author} {\bibfnamefont {K.~V.}\ \bibnamefont {Samokhin}},\ }\bibfield  {title} {\bibinfo {title} {Helical phases in superconductors},\ }\href@noop {} {\bibfield  {journal} {\bibinfo  {journal} {Journal of Experimental and Theoretical Physics}\ }\textbf {\bibinfo {volume} {78}},\ \bibinfo {pages} {401} (\bibinfo {year} {1994})},\ \bibinfo {note} {place: United States INIS Reference Number: 26032477}\BibitemShut {NoStop}%
\bibitem [{\citenamefont {Mineev}\ and\ \citenamefont {Samokhin}(2008)}]{MineevSamokhin08}%
  \BibitemOpen
  \bibfield  {author} {\bibinfo {author} {\bibfnamefont {V.~P.}\ \bibnamefont {Mineev}}\ and\ \bibinfo {author} {\bibfnamefont {K.~V.}\ \bibnamefont {Samokhin}},\ }\bibfield  {title} {\bibinfo {title} {Nonuniform states in noncentrosymmetric superconductors: {Derivation} of {Lifshitz} invariants from microscopic theory},\ }\href@noop {} {\bibfield  {journal} {\bibinfo  {journal} {Physical Review B}\ }\textbf {\bibinfo {volume} {78}},\ \bibinfo {pages} {144503} (\bibinfo {year} {2008})},\ \bibinfo {note} {publisher: American Physical Society}\BibitemShut {NoStop}%
\bibitem [{\citenamefont {Dimitrova}\ and\ \citenamefont {Feigel'man}(2007)}]{Dimitrova2007}%
  \BibitemOpen
  \bibfield  {author} {\bibinfo {author} {\bibfnamefont {O.}~\bibnamefont {Dimitrova}}\ and\ \bibinfo {author} {\bibfnamefont {M.~V.}\ \bibnamefont {Feigel'man}},\ }\bibfield  {title} {\bibinfo {title} {Theory of a two-dimensional superconductor with broken inversion symmetry},\ }\href {https://doi.org/10.1103/PhysRevB.76.014522} {\bibfield  {journal} {\bibinfo  {journal} {Phys. Rev. B}\ }\textbf {\bibinfo {volume} {76}},\ \bibinfo {pages} {014522} (\bibinfo {year} {2007})}\BibitemShut {NoStop}%
\bibitem [{\citenamefont {Kapustin}\ and\ \citenamefont {Radzihovsky}(2022)}]{KapustinRadzihovsky22}%
  \BibitemOpen
  \bibfield  {author} {\bibinfo {author} {\bibfnamefont {A.}~\bibnamefont {Kapustin}}\ and\ \bibinfo {author} {\bibfnamefont {L.}~\bibnamefont {Radzihovsky}},\ }\bibfield  {title} {\bibinfo {title} {Piezosuperconductivity: Novel effects in noncentrosymmetric superconductors},\ }\href {https://doi.org/10.1103/PhysRevB.105.134514} {\bibfield  {journal} {\bibinfo  {journal} {Phys. Rev. B}\ }\textbf {\bibinfo {volume} {105}},\ \bibinfo {pages} {134514} (\bibinfo {year} {2022})}\BibitemShut {NoStop}%
\bibitem [{\citenamefont {Kochan}\ \emph {et~al.}(2023)\citenamefont {Kochan}, \citenamefont {Costa}, \citenamefont {Zhumagulov},\ and\ \citenamefont {Zutic}}]{Kochan23}%
  \BibitemOpen
  \bibfield  {author} {\bibinfo {author} {\bibfnamefont {D.}~\bibnamefont {Kochan}}, \bibinfo {author} {\bibfnamefont {A.}~\bibnamefont {Costa}}, \bibinfo {author} {\bibfnamefont {I.}~\bibnamefont {Zhumagulov}},\ and\ \bibinfo {author} {\bibfnamefont {I.}~\bibnamefont {Zutic}},\ }\href@noop {} {\bibinfo {title} {Phenomenological {Theory} of the {Supercurrent} {Diode} {Effect}: {The} {Lifshitz} {Invariant}}} (\bibinfo {year} {2023}),\ \bibinfo {note} {arXiv:2303.11975 [cond-mat]}\BibitemShut {NoStop}%
\bibitem [{\citenamefont {Bauriedl}\ \emph {et~al.}(2022)\citenamefont {Bauriedl}, \citenamefont {B\"auml}, \citenamefont {Fuchs}, \citenamefont {Baumgartner}, \citenamefont {Paulik}, \citenamefont {Bauer}, \citenamefont {Lin}, \citenamefont {Lupton}, \citenamefont {Taniguchi}, \citenamefont {Watanabe}, \citenamefont {Strunk},\ and\ \citenamefont {Paradiso}}]{BauriedlParadiso22}%
  \BibitemOpen
  \bibfield  {author} {\bibinfo {author} {\bibfnamefont {L.}~\bibnamefont {Bauriedl}}, \bibinfo {author} {\bibfnamefont {C.}~\bibnamefont {B\"auml}}, \bibinfo {author} {\bibfnamefont {L.}~\bibnamefont {Fuchs}}, \bibinfo {author} {\bibfnamefont {C.}~\bibnamefont {Baumgartner}}, \bibinfo {author} {\bibfnamefont {N.}~\bibnamefont {Paulik}}, \bibinfo {author} {\bibfnamefont {J.~M.}\ \bibnamefont {Bauer}}, \bibinfo {author} {\bibfnamefont {K.-Q.}\ \bibnamefont {Lin}}, \bibinfo {author} {\bibfnamefont {J.~M.}\ \bibnamefont {Lupton}}, \bibinfo {author} {\bibfnamefont {T.}~\bibnamefont {Taniguchi}}, \bibinfo {author} {\bibfnamefont {K.}~\bibnamefont {Watanabe}}, \bibinfo {author} {\bibfnamefont {C.}~\bibnamefont {Strunk}},\ and\ \bibinfo {author} {\bibfnamefont {N.}~\bibnamefont {Paradiso}},\ }\bibfield  {title} {\bibinfo {title} {Supercurrent diode effect and magnetochiral anisotropy in few-layer {N}b{S}e$_2$},\ }\href@noop {} {\bibfield  {journal} {\bibinfo  {journal} {Nature Communications}\ }\textbf {\bibinfo
  {volume} {13}},\ \bibinfo {pages} {4266} (\bibinfo {year} {2022})},\ \bibinfo {note} {number: 1 Publisher: Nature Publishing Group}\BibitemShut {NoStop}%
\bibitem [{\citenamefont {Rikken}\ \emph {et~al.}(2001)\citenamefont {Rikken}, \citenamefont {F\"olling},\ and\ \citenamefont {Wyder}}]{Rikken:PRL2001}%
  \BibitemOpen
  \bibfield  {author} {\bibinfo {author} {\bibfnamefont {G.~L. J.~A.}\ \bibnamefont {Rikken}}, \bibinfo {author} {\bibfnamefont {J.}~\bibnamefont {F\"olling}},\ and\ \bibinfo {author} {\bibfnamefont {P.}~\bibnamefont {Wyder}},\ }\bibfield  {title} {\bibinfo {title} {Electrical magnetochiral anisotropy},\ }\href@noop {} {\bibfield  {journal} {\bibinfo  {journal} {Phys. Rev. Lett.}\ }\textbf {\bibinfo {volume} {87}},\ \bibinfo {pages} {236602} (\bibinfo {year} {2001})}\BibitemShut {NoStop}%
\bibitem [{\citenamefont {Wakatsuki}\ \emph {et~al.}(2017)\citenamefont {Wakatsuki}, \citenamefont {Saito}, \citenamefont {Hoshino}, \citenamefont {Itahashi}, \citenamefont {Ideue}, \citenamefont {Ezawa}, \citenamefont {Iwasa},\ and\ \citenamefont {Nagaosa}}]{WakatsukiNagaosa17}%
  \BibitemOpen
  \bibfield  {author} {\bibinfo {author} {\bibfnamefont {R.}~\bibnamefont {Wakatsuki}}, \bibinfo {author} {\bibfnamefont {Y.}~\bibnamefont {Saito}}, \bibinfo {author} {\bibfnamefont {S.}~\bibnamefont {Hoshino}}, \bibinfo {author} {\bibfnamefont {Y.~M.}\ \bibnamefont {Itahashi}}, \bibinfo {author} {\bibfnamefont {T.}~\bibnamefont {Ideue}}, \bibinfo {author} {\bibfnamefont {M.}~\bibnamefont {Ezawa}}, \bibinfo {author} {\bibfnamefont {Y.}~\bibnamefont {Iwasa}},\ and\ \bibinfo {author} {\bibfnamefont {N.}~\bibnamefont {Nagaosa}},\ }\bibfield  {title} {\bibinfo {title} {Nonreciprocal charge transport in noncentrosymmetric superconductors},\ }\href@noop {} {\bibfield  {journal} {\bibinfo  {journal} {Science Advances}\ }\textbf {\bibinfo {volume} {3}},\ \bibinfo {pages} {e1602390} (\bibinfo {year} {2017})},\ \bibinfo {note} {publisher: American Association for the Advancement of Science}\BibitemShut {NoStop}%
\bibitem [{\citenamefont {Legg}\ \emph {et~al.}(2022)\citenamefont {Legg}, \citenamefont {Loss},\ and\ \citenamefont {Klinovaja}}]{LeggLossKlinovaja22}%
  \BibitemOpen
  \bibfield  {author} {\bibinfo {author} {\bibfnamefont {H.~F.}\ \bibnamefont {Legg}}, \bibinfo {author} {\bibfnamefont {D.}~\bibnamefont {Loss}},\ and\ \bibinfo {author} {\bibfnamefont {J.}~\bibnamefont {Klinovaja}},\ }\bibfield  {title} {\bibinfo {title} {Superconducting diode effect due to magnetochiral anisotropy in topological insulators and {Rashba} nanowires},\ }\href@noop {} {\bibfield  {journal} {\bibinfo  {journal} {Physical Review B}\ }\textbf {\bibinfo {volume} {106}},\ \bibinfo {pages} {104501} (\bibinfo {year} {2022})},\ \bibinfo {note} {publisher: American Physical Society}\BibitemShut {NoStop}%
\bibitem [{\citenamefont {Yi}\ \emph {et~al.}(2022)\citenamefont {Yi}, \citenamefont {Hu}, \citenamefont {Wang}, \citenamefont {Xiao}, \citenamefont {Cai}, \citenamefont {Hickey}, \citenamefont {Dong}, \citenamefont {Zhao}, \citenamefont {Zhou}, \citenamefont {Zhang}, \citenamefont {Richardella}, \citenamefont {Alem}, \citenamefont {Robinson}, \citenamefont {Chan}, \citenamefont {Xu}, \citenamefont {Samarth}, \citenamefont {Liu},\ and\ \citenamefont {Chang}}]{Yi2022}%
  \BibitemOpen
  \bibfield  {author} {\bibinfo {author} {\bibfnamefont {H.}~\bibnamefont {Yi}}, \bibinfo {author} {\bibfnamefont {L.-H.}\ \bibnamefont {Hu}}, \bibinfo {author} {\bibfnamefont {Y.}~\bibnamefont {Wang}}, \bibinfo {author} {\bibfnamefont {R.}~\bibnamefont {Xiao}}, \bibinfo {author} {\bibfnamefont {J.}~\bibnamefont {Cai}}, \bibinfo {author} {\bibfnamefont {D.~R.}\ \bibnamefont {Hickey}}, \bibinfo {author} {\bibfnamefont {C.}~\bibnamefont {Dong}}, \bibinfo {author} {\bibfnamefont {Y.-F.}\ \bibnamefont {Zhao}}, \bibinfo {author} {\bibfnamefont {L.-J.}\ \bibnamefont {Zhou}}, \bibinfo {author} {\bibfnamefont {R.}~\bibnamefont {Zhang}}, \bibinfo {author} {\bibfnamefont {A.~R.}\ \bibnamefont {Richardella}}, \bibinfo {author} {\bibfnamefont {N.}~\bibnamefont {Alem}}, \bibinfo {author} {\bibfnamefont {J.~A.}\ \bibnamefont {Robinson}}, \bibinfo {author} {\bibfnamefont {M.~H.~W.}\ \bibnamefont {Chan}}, \bibinfo {author} {\bibfnamefont {X.}~\bibnamefont {Xu}}, \bibinfo {author} {\bibfnamefont {N.}~\bibnamefont {Samarth}},
  \bibinfo {author} {\bibfnamefont {C.-X.}\ \bibnamefont {Liu}},\ and\ \bibinfo {author} {\bibfnamefont {C.-Z.}\ \bibnamefont {Chang}},\ }\bibfield  {title} {\bibinfo {title} {{Crossover from Ising- to Rashba-type superconductivity in epitaxial Bi$_2$Se$_3$/monolayer NbSe$_2$ heterostructures}},\ }\href {https://doi.org/10.1038/s41563-022-01386-z} {\bibfield  {journal} {\bibinfo  {journal} {Nature Materials}\ }\textbf {\bibinfo {volume} {21}},\ \bibinfo {pages} {1366} (\bibinfo {year} {2022})}\BibitemShut {NoStop}%
\bibitem [{\citenamefont {Cho}\ \emph {et~al.}(2022)\citenamefont {Cho}, \citenamefont {Lyu}, \citenamefont {An}, \citenamefont {Han}, \citenamefont {Lo}, \citenamefont {Ng}, \citenamefont {Hu}, \citenamefont {Gao}, \citenamefont {Li}, \citenamefont {Huang}, \citenamefont {Wang}, \citenamefont {Schmalian},\ and\ \citenamefont {Lortz}}]{Cho2022}%
  \BibitemOpen
  \bibfield  {author} {\bibinfo {author} {\bibfnamefont {C.-w.}\ \bibnamefont {Cho}}, \bibinfo {author} {\bibfnamefont {J.}~\bibnamefont {Lyu}}, \bibinfo {author} {\bibfnamefont {L.}~\bibnamefont {An}}, \bibinfo {author} {\bibfnamefont {T.}~\bibnamefont {Han}}, \bibinfo {author} {\bibfnamefont {K.~T.}\ \bibnamefont {Lo}}, \bibinfo {author} {\bibfnamefont {C.~Y.}\ \bibnamefont {Ng}}, \bibinfo {author} {\bibfnamefont {J.}~\bibnamefont {Hu}}, \bibinfo {author} {\bibfnamefont {Y.}~\bibnamefont {Gao}}, \bibinfo {author} {\bibfnamefont {G.}~\bibnamefont {Li}}, \bibinfo {author} {\bibfnamefont {M.}~\bibnamefont {Huang}}, \bibinfo {author} {\bibfnamefont {N.}~\bibnamefont {Wang}}, \bibinfo {author} {\bibfnamefont {J.}~\bibnamefont {Schmalian}},\ and\ \bibinfo {author} {\bibfnamefont {R.}~\bibnamefont {Lortz}},\ }\bibfield  {title} {\bibinfo {title} {Nodal and nematic superconducting phases in ${\mathrm{nbse}}_{2}$ monolayers from competing superconducting channels},\ }\href
  {https://doi.org/10.1103/PhysRevLett.129.087002} {\bibfield  {journal} {\bibinfo  {journal} {Phys. Rev. Lett.}\ }\textbf {\bibinfo {volume} {129}},\ \bibinfo {pages} {087002} (\bibinfo {year} {2022})}\BibitemShut {NoStop}%
\bibitem [{\citenamefont {Attias}\ \emph {et~al.}(2024)\citenamefont {Attias}, \citenamefont {Michaeli},\ and\ \citenamefont {Khodas}}]{Attias2024}%
  \BibitemOpen
  \bibfield  {author} {\bibinfo {author} {\bibfnamefont {L.}~\bibnamefont {Attias}}, \bibinfo {author} {\bibfnamefont {K.}~\bibnamefont {Michaeli}},\ and\ \bibinfo {author} {\bibfnamefont {M.}~\bibnamefont {Khodas}},\ }\bibfield  {title} {\bibinfo {title} {{Planar Hall effect from superconducting fluctuations}},\ }\href {https://doi.org/10.1103/PhysRevB.110.014521} {\bibfield  {journal} {\bibinfo  {journal} {Phys. Rev. B}\ }\textbf {\bibinfo {volume} {110}},\ \bibinfo {pages} {014521} (\bibinfo {year} {2024})}\BibitemShut {NoStop}%
\bibitem [{\citenamefont {Cheng}\ \emph {et~al.}(2016)\citenamefont {Cheng}, \citenamefont {Sun}, \citenamefont {Chen}, \citenamefont {Fu},\ and\ \citenamefont {Meng}}]{Cheng_2018}%
  \BibitemOpen
  \bibfield  {author} {\bibinfo {author} {\bibfnamefont {C.}~\bibnamefont {Cheng}}, \bibinfo {author} {\bibfnamefont {J.-T.}\ \bibnamefont {Sun}}, \bibinfo {author} {\bibfnamefont {X.-R.}\ \bibnamefont {Chen}}, \bibinfo {author} {\bibfnamefont {H.-X.}\ \bibnamefont {Fu}},\ and\ \bibinfo {author} {\bibfnamefont {S.}~\bibnamefont {Meng}},\ }\bibfield  {title} {\bibinfo {title} {Nonlinear rashba spin splitting in transition metal dichalcogenide monolayers},\ }\href {https://doi.org/10.1039/C6NR04235J} {\bibfield  {journal} {\bibinfo  {journal} {Nanoscale}\ }\textbf {\bibinfo {volume} {8}},\ \bibinfo {pages} {17854} (\bibinfo {year} {2016})}\BibitemShut {NoStop}%
\bibitem [{\citenamefont {Xi}\ \emph {et~al.}(2015)\citenamefont {Xi}, \citenamefont {Zhao}, \citenamefont {Wang}, \citenamefont {Berger}, \citenamefont {Forró}, \citenamefont {Shan},\ and\ \citenamefont {Mak}}]{Xi2015}%
  \BibitemOpen
  \bibfield  {author} {\bibinfo {author} {\bibfnamefont {X.}~\bibnamefont {Xi}}, \bibinfo {author} {\bibfnamefont {L.}~\bibnamefont {Zhao}}, \bibinfo {author} {\bibfnamefont {Z.}~\bibnamefont {Wang}}, \bibinfo {author} {\bibfnamefont {H.}~\bibnamefont {Berger}}, \bibinfo {author} {\bibfnamefont {L.}~\bibnamefont {Forró}}, \bibinfo {author} {\bibfnamefont {J.}~\bibnamefont {Shan}},\ and\ \bibinfo {author} {\bibfnamefont {K.~F.}\ \bibnamefont {Mak}},\ }\bibfield  {title} {\bibinfo {title} {Strongly enhanced charge-density-wave order in monolayer nbse2},\ }\href {https://doi.org/10.1038/nnano.2015.143} {\bibfield  {journal} {\bibinfo  {journal} {Nature Nanotechnology}\ }\textbf {\bibinfo {volume} {10}},\ \bibinfo {pages} {765} (\bibinfo {year} {2015})}\BibitemShut {NoStop}%
\bibitem [{\citenamefont {Zhang}\ \emph {et~al.}(2014)\citenamefont {Zhang}, \citenamefont {Johnson}, \citenamefont {Hsu}, \citenamefont {Li},\ and\ \citenamefont {Shih}}]{Zhang2014}%
  \BibitemOpen
  \bibfield  {author} {\bibinfo {author} {\bibfnamefont {C.}~\bibnamefont {Zhang}}, \bibinfo {author} {\bibfnamefont {A.}~\bibnamefont {Johnson}}, \bibinfo {author} {\bibfnamefont {C.-L.}\ \bibnamefont {Hsu}}, \bibinfo {author} {\bibfnamefont {L.-J.}\ \bibnamefont {Li}},\ and\ \bibinfo {author} {\bibfnamefont {C.-K.}\ \bibnamefont {Shih}},\ }\bibfield  {title} {\bibinfo {title} {Direct imaging of band profile in single layer mos2 on graphite: Quasiparticle energy gap, metallic edge states, and edge band bending},\ }\href {https://doi.org/10.1021/nl501133c} {\bibfield  {journal} {\bibinfo  {journal} {Nano Letters}\ }\textbf {\bibinfo {volume} {14}},\ \bibinfo {pages} {2443} (\bibinfo {year} {2014})},\ \bibinfo {note} {pMID: 24783945},\ \Eprint {https://arxiv.org/abs/https://doi.org/10.1021/nl501133c} {https://doi.org/10.1021/nl501133c} \BibitemShut {NoStop}%
\bibitem [{\citenamefont {Korm\'anyos}\ \emph {et~al.}(2015)\citenamefont {Korm\'anyos}, \citenamefont {Burkard}, \citenamefont {Gmitra}, \citenamefont {Fabian}, \citenamefont {Z\'olyomi}, \citenamefont {Drummond},\ and\ \citenamefont {Fal'ko}}]{Kormanyos_2015}%
  \BibitemOpen
  \bibfield  {author} {\bibinfo {author} {\bibfnamefont {A.}~\bibnamefont {Korm\'anyos}}, \bibinfo {author} {\bibfnamefont {G.}~\bibnamefont {Burkard}}, \bibinfo {author} {\bibfnamefont {M.}~\bibnamefont {Gmitra}}, \bibinfo {author} {\bibfnamefont {J.}~\bibnamefont {Fabian}}, \bibinfo {author} {\bibfnamefont {V.}~\bibnamefont {Z\'olyomi}}, \bibinfo {author} {\bibfnamefont {N.~D.}\ \bibnamefont {Drummond}},\ and\ \bibinfo {author} {\bibfnamefont {V.}~\bibnamefont {Fal'ko}},\ }\bibfield  {title} {\bibinfo {title} {k·p theory for two-dimensional transition metal dichalcogenide semiconductors},\ }\href {https://doi.org/10.1088/2053-1583/2/2/022001} {\bibfield  {journal} {\bibinfo  {journal} {2D Materials}\ }\textbf {\bibinfo {volume} {2}},\ \bibinfo {pages} {022001} (\bibinfo {year} {2015})}\BibitemShut {NoStop}%
\bibitem [{\citenamefont {Tang}\ \emph {et~al.}(2021{\natexlab{a}})\citenamefont {Tang}, \citenamefont {Klees}, \citenamefont {Bruder},\ and\ \citenamefont {Belzig}}]{Tang2021}%
  \BibitemOpen
  \bibfield  {author} {\bibinfo {author} {\bibfnamefont {G.}~\bibnamefont {Tang}}, \bibinfo {author} {\bibfnamefont {R.~L.}\ \bibnamefont {Klees}}, \bibinfo {author} {\bibfnamefont {C.}~\bibnamefont {Bruder}},\ and\ \bibinfo {author} {\bibfnamefont {W.}~\bibnamefont {Belzig}},\ }\bibfield  {title} {\bibinfo {title} {Controlling charge and spin transport in an ising-superconductor josephson junction},\ }\href {https://doi.org/10.1103/PhysRevB.104.L241413} {\bibfield  {journal} {\bibinfo  {journal} {Phys. Rev. B}\ }\textbf {\bibinfo {volume} {104}},\ \bibinfo {pages} {L241413} (\bibinfo {year} {2021}{\natexlab{a}})}\BibitemShut {NoStop}%
\bibitem [{\citenamefont {M\"ockli}\ and\ \citenamefont {Khodas}(2019)}]{Mockli2019}%
  \BibitemOpen
  \bibfield  {author} {\bibinfo {author} {\bibfnamefont {D.}~\bibnamefont {M\"ockli}}\ and\ \bibinfo {author} {\bibfnamefont {M.}~\bibnamefont {Khodas}},\ }\bibfield  {title} {\bibinfo {title} {Magnetic-field induced $s+\mathit{if}$ pairing in ising superconductors},\ }\href {https://doi.org/10.1103/PhysRevB.99.180505} {\bibfield  {journal} {\bibinfo  {journal} {Phys. Rev. B}\ }\textbf {\bibinfo {volume} {99}},\ \bibinfo {pages} {180505} (\bibinfo {year} {2019})}\BibitemShut {NoStop}%
\bibitem [{\citenamefont {Tang}\ \emph {et~al.}(2021{\natexlab{b}})\citenamefont {Tang}, \citenamefont {Bruder},\ and\ \citenamefont {Belzig}}]{Tang2021a}%
  \BibitemOpen
  \bibfield  {author} {\bibinfo {author} {\bibfnamefont {G.}~\bibnamefont {Tang}}, \bibinfo {author} {\bibfnamefont {C.}~\bibnamefont {Bruder}},\ and\ \bibinfo {author} {\bibfnamefont {W.}~\bibnamefont {Belzig}},\ }\bibfield  {title} {\bibinfo {title} {Magnetic field-induced ``mirage'' gap in an ising superconductor},\ }\href {https://doi.org/10.1103/PhysRevLett.126.237001} {\bibfield  {journal} {\bibinfo  {journal} {Phys. Rev. Lett.}\ }\textbf {\bibinfo {volume} {126}},\ \bibinfo {pages} {237001} (\bibinfo {year} {2021}{\natexlab{b}})}\BibitemShut {NoStop}%
\end{thebibliography}


\end{document}